\documentclass[twocolumn]{aastex631}
\usepackage{graphicx}	
\usepackage[caption=false]{subfig}
\usepackage{rotating}

\usepackage{amssymb}
\usepackage{color}
\usepackage{amsmath}
\usepackage{longtable}
\usepackage{threeparttablex}

\usepackage{natbib}
\newcommand{\ie}{\textit{i.e.,}}
\newcommand{\eg}{\textit{e.g.,}}

\newcommand{\logM}{$\log(M_*/\rm{M}_\odot)$}
\newcommand{\zphot}{$z_\textrm{phot}$}
\newcommand{\zspec}{$z_\textrm{spec}$}
\newcommand{\UVJ}{\textit{UVJ}}

\newcommand{\VJc}{\textit{(V-J)}}
\newcommand{\Dfour}{$D_n$(4000)}

\newcommand{\OII}{\hbox{{\rm [O}\kern 0.1em{\sc ii}{\rm ]$\lambda\lambda3726,3729$}}}
\newcommand{\OIII}{\hbox{{\rm [O}\kern 0.1em{\sc iii}{\rm ]$\lambda\lambda4959,5007$}}}

\newcommand{\Hbeta}{$\rm{H}\beta$}
\newcommand{\Halpha}{$\rm{H}\alpha$}
\newcommand{\NII}{[N~{\rm \scriptsize II}]$\lambda\lambda6548,6584$}
\newcommand{\SII}{[S~{\rm \scriptsize II}]$\lambda\lambda6718,6733$}
\newcommand{\SIII}{[S~{\rm \scriptsize III}]$\lambda9531$}
\newcommand{\CI}{[C~{\rm \scriptsize I}]$\lambda9850$}
\newcommand{\Bagpipes}{B{\sc agpipes}}
\newcommand{\sumg}{S-UMG}

\begin{document}

\title{MAGAZ3NE: Massive, Extremely Dusty Galaxies at $z\sim2$ Lead to Photometric Overestimation of Number Densities of the Most Massive Galaxies at $3<z<4$\footnote{The spectra presented herein were obtained at the W. M. Keck Observatory, which is operated as a scientific partnership among the California Institute of Technology, the University of California and the National Aeronautics and Space Administration. The Observatory was made possible by the generous financial support of the W. M. Keck Foundation.} }
\shorttitle{Massive Galaxy Overestimation}
\shortauthors{B. Forrest, et al.}

\correspondingauthor{Ben Forrest}
\email{bforrest@ucdavis.edu}

\author[0000-0001-6003-0541]{Ben Forrest}
	\affiliation{Department of Physics and Astronomy, University of California Davis, One Shields Avenue, Davis, CA, 95616, USA}
\author[0000-0003-1371-6019]{M. C. Cooper}
	\affiliation{Department of Physics and Astronomy, University of California, Irvine, 4129 Frederick Reines Hall, Irvine, CA 92697, USA}
\author[0000-0002-9330-9108]{Adam Muzzin}
	\affiliation{Department of Physics and Astronomy, York University, 4700, Keele Street, Toronto, ON MJ3 1P3, Canada}
\author[0000-0002-6572-7089]{Gillian Wilson}
	\affiliation{Department of Physics, University of California Merced, 5200 North Lake Rd., Merced, CA 95343, USA}
\author[0000-0001-9002-3502]{Danilo Marchesini}
	\affiliation{Department of Physics and Astronomy, Tufts University, 574 Boston Avenue, Medford, MA 02155, USA}
\author[0000-0002-2446-8770]{Ian McConachie}
	\affiliation{Department of Physics and Astronomy, University of California, Riverside, 900 University Avenue, Riverside, CA 92521, USA}
\author[0000-0003-0408-9850]{Percy Gomez}
	\affiliation{W.M. Keck Observatory, 65-1120 Mamalahoa Hwy., Kamuela, HI 96743, USA}
\author[0000-0002-8053-8040]{Marianna Annunziatella}
	\affiliation{Centro de Astrobiolog\'ia (CSIC-INTA), Ctra de Torrej\'on a Ajalvir, km 4, E-28850 Torrej\'on de Ardoz, Madrid, Spain}
\author[0000-0002-7248-1566]{Z. Cemile Marsan}
	\affiliation{Department of Physics and Astronomy, York University, 4700, Keele Street, Toronto, ON MJ3 1P3, Canada}
	
\author{Joey Braspenning}
	\affiliation{Leiden Observatory, Leiden University, PO Box 9513, 2300 RA Leiden, the Netherlands}
\author[0000-0003-2144-2943]{Wenjun Chang}
	\affiliation{Department of Physics and Astronomy, University of California, Riverside, 900 University Avenue, Riverside, CA 92521, USA}
\author{Gabriella de Lucia}
	\affiliation{INAF - Astronomical Observatory of Trieste, via G.B. Tiepolo 11, Trieste, Italy}
	\affiliation{IFPU - Institute for Fundamental Physics of the Universe, via Beirut 2, 34151, Trieste, Italy}
\author[0000-0003-4744-0188]{Fabio Fontanot}
	\affiliation{INAF - Astronomical Observatory of Trieste, via G.B. Tiepolo 11, Trieste, Italy}
	\affiliation{IFPU - Institute for Fundamental Physics of the Universe, via Beirut 2, 34151, Trieste, Italy}
\author[0000-0002-3301-3321]{Michaela Hirschmann}
	\affiliation{Institute of Physics, Lab for galaxy evolution, EPFL, Observatoire de Sauverny, Chemin Pegasi 51, 1290 Switzerland}
	\affiliation{INAF - Astronomical Observatory of Trieste, via G.B. Tiepolo 11, Trieste, Italy}
\author[0000-0001-8421-5890]{Dylan Nelson}
	\affiliation{Universit\"at Heidelberg, Institut f\"ur Theoretische Astrophysik, ZAH, Albert-Ueberle-Str. 2, 69120 Heidelberg, Germany}
\author[0000-0003-1065-9274]{Annalisa Pillepich}
	\affiliation{Max-Planck-Institut f{\"u}r Astronomie, K{\"o}nigstuhl 17, 69117 Heidelberg, Germany}
\author[0000-0002-0668-5560]{Joop Schaye}
	\affiliation{Leiden Observatory, Leiden University, PO Box 9513, 2300 RA Leiden, the Netherlands}
\author[0000-0001-8169-7249]{Stephanie M. Urbano Stawinski}
	\affiliation{Department of Physics and Astronomy, University of California, Irvine, 4129 Frederick Reines Hall, Irvine, CA 92697, USA}
\author{Mauro Stefanon}
	\affiliation{Departament d'Astronomia i Astrof\`isica, Universitat de Val\`encia, C. Dr. Moliner 50, E-46100 Burjassot, Val\`encia, Spain}
	\affiliation{Unidad Asociada CSIC ``Grupo de Astrof\'isica Extragal\'actica y Cosmolog\'ia" (Instituto de F\'isica de Cantabria - Universitat de Val\`encia)}
\author[0000-0003-3864-068X]{Lizhi Xie}
	\affiliation{Tianjin Normal University, Binshuixidao 393, Xiqing, 300387, Tianjin, China}

\keywords{galaxies: high-redshift -- galaxies: evolution}

\begin{abstract}

We present rest-frame optical spectra from Keck/MOSFIRE and Keck/NIRES of 16 candidate ultramassive galaxies targeted as part of the Massive Ancient Galaxies at $z>3$ Near-Infrared (MAGAZ3NE) Survey.
These candidates were selected to have photometric redshifts \mbox{$3\lesssim$~\zphot~$<4$}, photometric stellar masses \mbox{\logM~$>11.7$}, and well-sampled photometric spectral energy distributions (SEDs) from the \mbox{UltraVISTA} and VIDEO surveys.
In contrast to previous spectroscopic observations of blue star-forming and post-starburst ultramassive galaxies, candidates in this sample have very red SEDs
implying significant dust attenuation, old stellar ages, and/or active galactic nuclei (AGN).
Of these galaxies, eight are revealed to be heavily dust-obscured $2.0<z<2.7$ galaxies with strong emission lines, some showing broad features indicative of AGN, three are Type I AGN hosts at $z>3$, one is a $z\sim1.2$ dusty galaxy, and four galaxies do not have a confirmed spectroscopic redshift.
In fact, none of the sample has \mbox{$\lvert$~\zspec-\zphot~$\rvert <0.5$}, suggesting difficulties for photometric redshift programs in fitting similarly red SEDs.
The prevalence of these red interloper galaxies suggests that the number densities of high-mass galaxies are overestimated at $z\gtrsim3$ in large photometric surveys, helping to resolve the `impossibly early galaxy problem' and leading to much better agreement with cosmological galaxy simulations.
A more complete spectroscopic survey of ultramassive galaxies is required to pin down the uncertainties on their number densities in the early universe.

\end{abstract}

%%%%%%%%%%%%%%%%%%%%%%%%%%%%%%%%%%%%%%%%%%%%%%%%%%

\section{Introduction}

Measuring the change in number density of galaxies versus stellar mass as a function of cosmic time provides insights into the build-up of stellar mass and, when considering star-forming and quiescent galaxies separately, the timescales over which galaxies quench their star formation.
This galaxy stellar mass function (GSMF) has therefore become a standard tool for studying galaxy evolution.

The GSMF has been measured using optical surveys across a range of redshifts from the local Universe \citep[\eg][]{Cole2001, Bell2003} out to $z\sim1.5$ \citep[\eg][]{Bundy2006, Borch2006}.
At these redshifts, the evolution of the GSMF appears to be small.
The most massive galaxies are already in place and are typically only growing via minor mergers (\ie\ slowly), while star formation rates in less massive galaxies are lower than at earlier epochs \citep[\eg][]{Kawinwanichakij2020a}.

Wide and deep near-infrared imaging surveys have allowed for GSMF analyses out to considerably higher redshifts and have also probed down to lower stellar masses \citep[\eg][]{Marchesini2009, Muzzin2013, Ilbert2013, Tomczak2014, Stefanon2015, Davidzon2017, Sherman2020a, McLeod2020, Marsan2022, Weaver2023a}.
These studies have shown that the GSMF evolves substantially in the first 4 Gyr as galaxies form new stars rapidly and the first galaxies begin quenching.
Intriguingly, the number densities of the most massive galaxies (\logM~$\gtrsim11.5$) do not appear to change much, if at all, after this early epoch \citep{Marchesini2010, Kawinwanichakij2020a}, implying that nearly all of these objects which exist today have built the majority of their stellar mass in the first $\sim3$~Gyr of cosmic time and then ceased stellar mass growth almost completely.
The number densities of these massive galaxies at high redshifts are significantly higher than those predicted by most simulations, and 
are close to, if not in excess of, the limits of hierarchical galaxy assembly and $\Lambda$CDM \citep[\eg][]{Antwi-Danso2023b, Glazebrook2024}.
This has been termed the `impossibly early galaxy problem' \citep{Steinhardt2016}, though extremely efficient star-formation in early, massive halos \citep[\eg][]{Stefanon2021a, Weaver2023a} or a varying initial mass function \citep[\eg][]{vanDokkum2012, Mendel2020} may help resolve the issue.

Spectroscopic confirmation of these most massive galaxies at early times with accurate measurements of their stellar mass is required to understand whether changes to the current paradigm of galaxy evolution are necessary.
In the last decade, samples of ultra-massive galaxies (UMGs; \logM~$\gtrsim11$) at $3<z<4$ have been confirmed with optical/near-infrared spectroscopy to the point where there can be no doubt that such galaxies exist in substantial numbers \citep[\eg][]{Marsan2015, Glazebrook2017, Schreiber2018b, Tanaka2019, Valentino2020, Forrest2020a, Forrest2020b, Saracco2020, Antwi-Danso2023b}.
Several have also been spectroscopically confirmed with longer wavelength observations \citep[\eg][]{Jones2021}.
Importantly however, the difficulty of confirming these objects means that the brightest candidates are typically selected for spectroscopic followup.
This biases the sample toward confirmation of blue star-forming and post-starburst galaxies at these epochs, which have rest-frame $V$-band magnitudes $\sim10\times$ higher than older quiescent galaxies with similar stellar masses \citep[\eg][]{Bruzual2003}.
Additionally, none of these UMGs confirmed with optical/near-infrared spectroscopy are highly dust-obscured galaxies, which photometric surveys suggest contribute significantly to the population of high-mass galaxies at these redshifts \citep[\eg][]{Martis2016, Marsan2022, Long2023}, though see also \citet{Dunlop2007}.
Such galaxies are extremely difficult to confirm spectroscopically in the optical/near-infrared even using the largest ground-based telescopes as dust obscuration reduces the strength of emission lines and absorption features in the rest-frame optical typically used for confirmation. 

We have undertaken an ambitious campaign to spectroscopically confirm candidate UMGs with \logM~$>11.7$ and $z>3$ (herein termed super-ultramassive galaxies; \sumg s) with Keck/MOSFIRE \citep{McLean2010, McLean2012}, many of which the photometry suggests may be dust-obscured.
In this work we describe the photometric sample, target selection, spectroscopic observations, and data reduction in Section~\ref{Sec:Data}, and analysis of the spectroscopy in Section~\ref{Sec:AnSpec}.
This is followed by a comparison of \sumg\ candidate objects in the COSMOS field from different catalogs in Section~\ref{Sec:AnComp} and discussion of their number densities in Section~\ref{Sec:Disc}.
Throughout this work we assume a $\Lambda$CDM cosmology with $H_0=70$~km~s$^{-1}$~Mpc$^{-1}$, $\Omega_M=0.3$, and $\Omega_\Lambda=0.7$.
We also adopt the AB magnitude system \citep{Oke1983}.

\section{Data} \label{Sec:Data}

\subsection{Photometric Catalogs}

This work builds on previous observations from the MAGAZ3NE survey, which spectroscopically confirmed the largest sample to date of UMGs at $3<z<4$ using Keck/MOSFIRE \citep{Forrest2020b}.
Previous targets were almost entirely either blue star-forming or post-starburst galaxies and showed excellent agreement between photometric and spectroscopic redshifts and stellar masses.
This population also showed large velocity dispersions \citep{Forrest2022} and the residence of at least some of the UMGs in confirmed overdense structures \citep[][McConachie et al. 2024, submitted]{Shen2021, McConachie2022}.
In this work we target a sample which is considerably redder in color and contains the most massive candidate galaxies at $z>3$. 

As in previous MAGAZ3NE papers, we select targets for spectroscopic follow-up from wide and deep near-infrared selected photometric catalogs.
Two catalogs in the COSMOS field \citep{Scoville2007} based on deep near-infrared $Y$-, $J$-, $H$-, and $K_s$-band data from the UltraVISTA survey \citep{McCracken2012} are considered.
The first catalog \citep{Muzzin2013a} uses the UltraVISTA data release 1 over 1.62~deg$^2$ in addition to ground-based optical data \citep[Subaru/SuprimeCam and CFHT/MegaCam;][]{Taniguchi2007, Capak2007} and space-based ultraviolet \citep[GALEX;][]{Martin2005} and near-infrared \citep[Spitzer/ IRAC and Spitzer/MIPS;][]{Sanders2007} data.
Photometry is based on a $K_s$-band detection image with a 90\% completeness of $K_s=23.4$~mag and 30 photometric bandpasses.
The second catalog \citep[][A. Muzzin, private communication]{Marsan2022} is based on UltraVISTA data release 3, which includes deeper observations ($K_s=24.5$~mag) in strips totaling 0.84~deg$^2$.
Additional observations from Spitzer/IRAC \citep{Mehta2018, Caputi2017, Ashby2018} increased the depth in the 3.6~$\mu$m and 4.5~$\mu$m bands by $>1$~mag, while additional data from Subaru/Hyper Suprime-Cam \citep{Aihara2019}, CFHTLS-Deep, and the NEWFIRM Medium Band Survey \citep{Whitaker2011} were also added for up to 49 photometric bandpasses per object.

Two catalogs based on $Z$-, $Y$-, $J$-, $H$-, and $K_s$-band imaging from the VISTA VIDEO survey \citep{Jarvis2013} are also used (M. Annunziatella, private communication).
These catalogs, one in the \textit{XMM}-LSS (XMM) field and one in the \textit{Chandra} Deep Field South (CDFS), were constructed following a similar process to the UltraVISTA catalogs.
The VIDEO observations were combined with observations from the Dark Energy Survey \citep{Abbott2018}, CFHT-Deep/Wide Legacy Survey \citep{Gwyn2012}, VOICE Survey \citep{Vaccari2016}, and Hyper-SuprimeCam \citep{Hayashi2018}.
Deep IRAC observations from the SERVS \citep{Mauduit2012} and DeepDrill \citep{Lacy2021} surveys were also included for a total of 22 photometric bandpasses.
The XMM catalog covers 5.1~deg$^2$ while the CDFS catalog covers 3.3~deg$^2$.
Both have $K_s$-band detection images with 90\% completeness limits of $K_s\sim23.0$~mag. 

Galaxies in all catalogs have photometric redshift probability distributions from EaZY \citep{Brammer2008} and stellar population properties derived using FAST \citep{Kriek2009} assuming a \citet{Chabrier2003} IMF.

\subsection{Target Selection}

We apply a photometric stellar mass cut of \logM~$>11.7$ and a photometric redshift cut of $3<$~\zphot~$<4$ to initially select galaxies in these catalogs.
Each galaxy is visually inspected in an attempt to verify that it is a single object without contamination from a neighbor nearby in projection and that the best fit spectral energy distribution (SED) model is a reasonable fit to the photometry.
While a magnitude cut was not explicitly made for target selection, brighter targets generally had better fit SEDs and were more likely to be observed.
Only two observed candidates, both in COSMOS, had $K$-band magnitudes fainter than the cut of $K_s<21.7$ used in previous MAGAZ3NE works.

\begin{figure*}
	\includegraphics[width=\textwidth, trim=0in 0.4in 0in 0in]{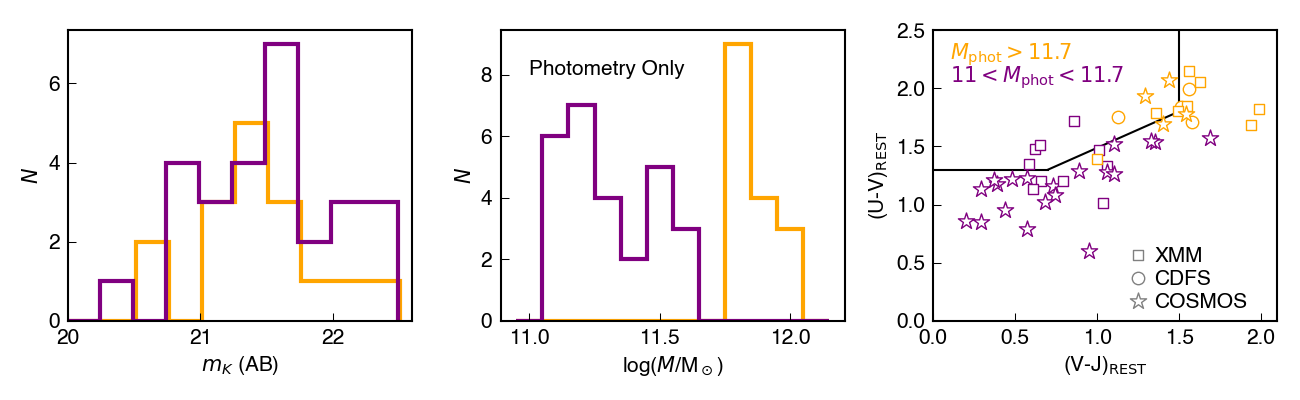}
    \caption{The $K$-band magnitudes (left), stellar masses derived from photometry (center), and UVJ rest-frame colors derived from photometry (right) of the MAGAZ3NE UMGs ($11.0<$~\logM~$<11.7$, purple) and \sumg s (\logM~$>11.7$, orange) candidates. The shape of a symbol indicates the field, either VIDEO-XMM (square), VIDEO-CDFS (circle), or COSMOS-UltraVISTA (star).}
    \label{fig:samp}
\end{figure*}

This results in a sample of 34 high-quality targets of which 16 were spectroscopically observed - 8/16 in XMM, 4/8 in CDFS, and 4/10 in COSMOS.
A comparison of the stellar masses and rest-frame colors from this sample and those targets from previous MAGAZ3NE work is shown in Figure~\ref{fig:samp}.

\subsection{Spectroscopy}

\subsubsection{Observations}

Fifteen targets were followed-up using the MOSFIRE instrument on the Keck I telescope.
These were initially targeted in the $K$-band, where the potentially strong \OIII\ and \Hbeta\ emission lines lie across $3<z<3.8$.
As with previous MAGAZ3NE follow-up, on-the-fly data reduction was performed to look for emission lines during the course of observations.
If a redshift was considered confirmed, we moved on to the next target.
As a result, the exposure times on each galaxy are non-uniform.
When a target showed a single emission line in the $K$-band whose nature could not reliably be determined, or when only continuum was detected, we followed-up in $H$-band.

We also targeted one \sumg\ candidate (XMM-VID1-2433) with the NIRES instrument \citep{JWilson2004} on Keck II.
While NIRES is less sensitive than MOSFIRE in the $K$-band and targets a single object at a time, it has the advantage of obtaining data in the \mbox{$Y$-, $J$-, $H$-,} and $K$-bandpasses simultaneously.
The exposure times and conditions for \sumg\ candidates are summarized in Table~\ref{tab:obs2}.
A more complete table of MAGAZ3NE targets is provided in the Appendix.

\subsubsection{Data Reduction}

In contrast to previous MAGAZ3NE papers which used the Keck-supported Data Reduction Pipeline (DRP) to reduce the data, for this work we reduced the data using the MOSDEF 2D data reduction pipeline \citep{Kriek2015,Kriek2024}.
The main advantage of this reduction method over the DRP is the use of a bright star on a science slit to account for slight pointing offsets between exposures and weight individual frames by throughput and seeing, which results in a higher SNR 2D-spectrum.
We also find a slightly more accurate accounting of spectral errors.
Previous MAGAZ3NE MOSFIRE data, which were initially reduced using the MOSFIRE DRP, have been re-reduced using the MOSDEF pipeline as well.
We note that in no case does this choice of reduction software significantly affect the redshift of an UMG in our sample.

Data from Keck/NIRES were reduced using v1.14 of Pypeit \citep{Prochaska2020} and a telluric correction was calculated using Molecfit \citep{Smette2015, Kausch2015}.

A custom Python code was used to perform optimal extraction of the targets \citep{Horne1986}.
When one or more emission lines were present, the shape of the optimal aperture was calculated around the strongest emission line.

%------------------------------------------
%------------------------------------------

%------------------------------------------
\begin{table*}
	\centering
	\caption{Spectral observations of \sumg\ candidates. All targets were observed with Keck/MOSFIRE with the exception of XMM-VID1-2433, which was observed with Keck/NIRES.}
	\label{tab:obs2}
	\begin{tabular}{lcrcl}
		\hline
		Galaxy & $m_{\rm K}$  & $t_{\rm exp,K} (m)$ & $t_{\rm exp,H} (m)$ & Avg. Seeing (K;\arcsec)  \\
		\hline			
		COS-DR1-70455	& 21.85	& 312	& --	& 0.58  \\
		COS-DR1-79837	& 21.10	& 120	& 216 & 0.74  \\
		COS-DR1-209435	& 22.03	& 252	& --	& 0.68  \\
		COS-DR1-233254	& 21.10	& 69		& --	& 0.77  \\
		\hline
		CDFS-VID1-2420	& 22.39	& 132	& --	& 0.74  \\
		CDFS-VID1-3091	& 20.59	& 114	& 40	& 0.60  \\
		CDFS-VID1-3536	& 21.44	& 204	& -- 	& 0.67  \\
		CDFS-VID2-684	& 21.67	& 210	& --	& 1.0	 \\
		\hline
		XMM-VID1-1415	& 21.34	& 66		& --	& 0.70  \\
		XMM-VID1-2433*	& 20.66	& 260	& = (260) & 0.80 \\
		XMM-VID2-827		& 21.41	& 72		& --	& 0.66  \\
		XMM-VID2-2379	& 21.71	& 96		& --	& 0.64  \\
		XMM-VID3-657		& 21.66	& 360	& 64	& 0.69  \\
		XMM-VID3-1787	& 21.15	& 114	& --	& 0.60  \\
		XMM-VID3-3517	& 21.42	& 296	& --	& 0.85  \\
		XMM-VID3-3941	& 21.33	& 180	& 88	& 1.1	  \\
	\end{tabular}
\end{table*}
%------------------------------------------

\section{Analysis of Spectra}\label{Sec:AnSpec}

\subsection{Spectroscopic Redshifts}

\begin{figure*}
	\includegraphics[width=\textwidth, trim=0in 4.4in 3.75in 0in]{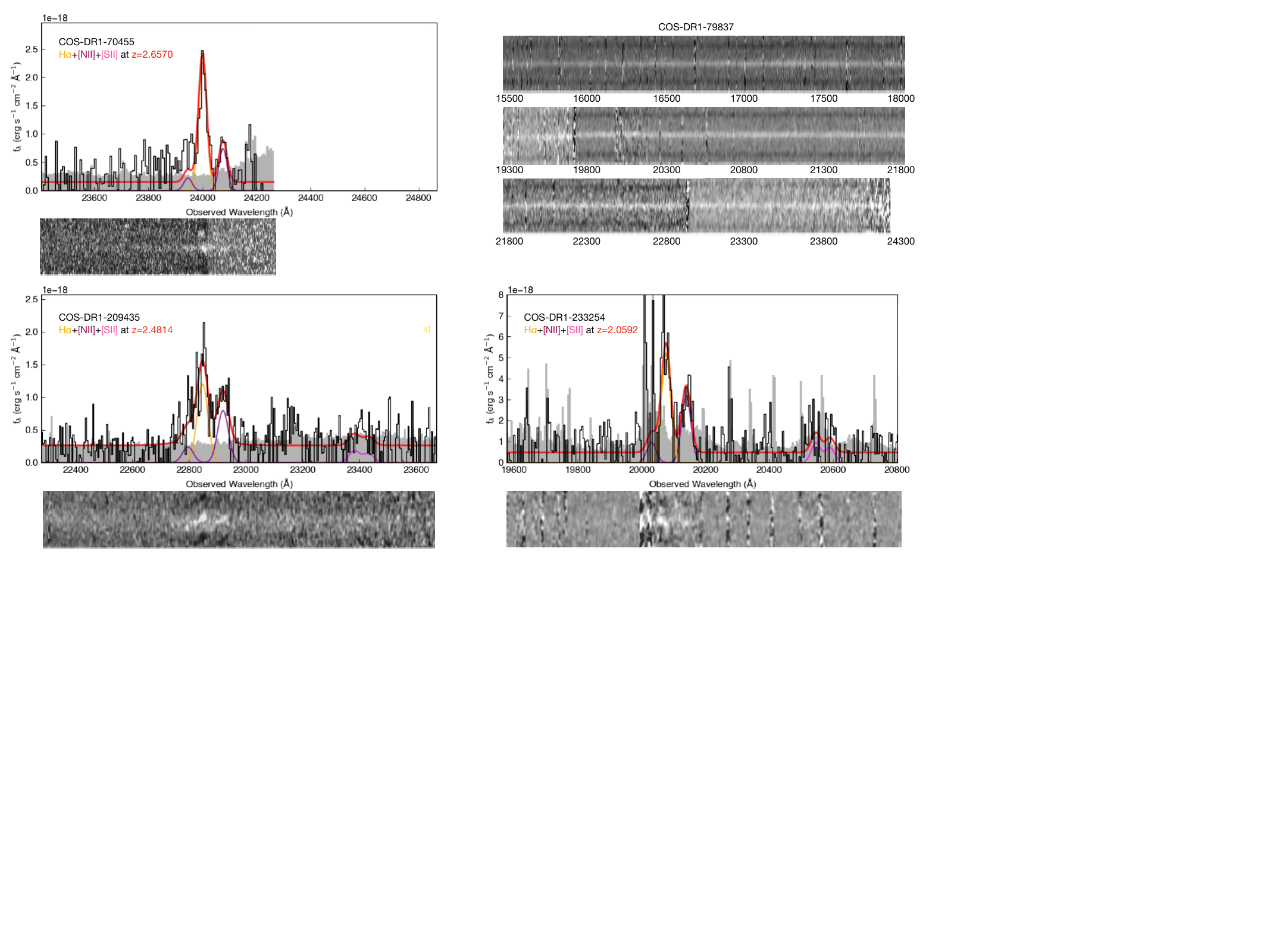}
    \caption{Line complex fitting for obtaining spectroscopic redshifts of candidate \sumg s in the COSMOS field. For objects with confirmed redshifts, the region of the 1D and 2D spectra with the confirming lines and fitted model are shown. The overall model is shown in red, while individual line components are shown in different colors (\Halpha\ in orange, \NII\ in purple, and \SII\ in magenta). For objects without confirmed redshifts (\eg\ COS-DR1-79837, top right), the entire wavelength range probed by $H$- and $K$-band spectroscopy is shown. Some targets were observed with multiple masks which were sensitive to different wavelengths; in such cases regions covered by fewer than the total number of masks are displayed as a lighter shade of gray. Similar plots for the targeted \sumg\ candidates in the VIDEO fields are in the Appendix.}
    \label{fig:speclines}
\end{figure*}

Of the 16 observed candidate \sumg s, 12 show obvious emission lines; 8 of these galaxies clearly display the \Halpha +\NII\ lines (and therefore at $2.0\lesssim z \lesssim2.7$), while 3 show the \OIII+\Hbeta\ lines at $z>3$, and one target exhibits \SIII\ and \CI\ at $z\sim1.2$.
These are fit with a multi-gaussian model, with the \NII\ doublet, \SII\ doublet, and \OIII\ doublet line ratios fixed.
The velocity width for the narrow component of all lines is fixed, but an additional broad line component is also allowed.
Figure~\ref{fig:speclines} shows observed spectra and the model of the emission features in the COSMOS targets, which leave no uncertainty as to the galaxy redshift modulo small observational uncertainties (XMM and CDFS targets are similarly displayed in the Appendix).
 
The spectra of 3 of the remaining candidate \sumg s (XMM-VID3-657, XMM-VID3-1787, and COS-DR1-79837) have clear continuum detections but no strong emission features or absorption features that could be used to determine a redshift, while the final candidate, XMM-VID1-1415, has only a very faint continuum detection.
This implies that either the redshift of each galaxy is such that any strong emission lines fall out of the wavelength range to which the spectroscopic observations are sensitive, or that the galaxies do not have strong observable emission lines, either due to passive nature or high levels of dust-obscuration. 
Based on the photometric redshift probability distributions, the probability of these galaxies being located at a redshift where strong emission lines would not be observable for these targets is $\lesssim10\%$.
Fits to the combined photometry and spectroscopy suggest this may in fact be the case for both XMM-VID3-1787 and XMM-VID1-1415.
XMM-VID3-657 has a low signal-to-noise detection at $\sim2.36~\mu$m, which may correspond to \Hbeta\ at $z=3.85$, while a continuum break may be present in COS-DR1-79837.
Deeper near-infrared spectroscopic observations from the ground are unlikely to yield additional information, though observations with \textit{JWST} would likely result in confirmed redshifts.

\subsection{Remodeling the Stellar Populations}

The UltraVISTA and VIDEO catalogs use FAST to estimate stellar masses from photometry, which does not natively include emission lines in the models.
While it has been known for some time that emission lines can contribute a significant fraction of flux in broadband measurements \citep[\eg][]{Kriek2011, Stark2013, Smit2015, Salmon2015, Forrest2018}, particularly for low-mass, blue, star-forming galaxies, recent results with JWST have shown that this is quite common at high redshifts \citep[\eg][]{Withers2023, Sarrouh2024}.
Indeed, in this work we also see significant line flux contribution in very red, massive galaxies.

Using the spectroscopic redshift, we remodel the stellar populations of the galaxy sample using FAST++ \citep{Schreiber2018a} and employing the \citet{Bruzual2003} stellar population sythesis models.
FAST++ has the advantage of allowing for flexible star formation history parameterizations and including both photometric and spectroscopic data, though models still do not include emission lines.
We therefore perform a correction of the photometry due to observed emission features as in \citet{Forrest2020b} with the line models described above and mask the emission line wavelengths in the spectra, but note that the possibility of other strong lines at spectrally unobserved wavelengths may additionally affect the modeling.
The percentage contribution of flux from emission lines in the $K$-band for the \sumg\ candidate sample is $12.2\pm3.5\%$.
The best fit models overlaid on the observed photometry and spectra for objects with confirmed spectroscopic redshifts are shown in Figure~\ref{fig:specfit1}, while those without confirmed spectroscopic redshifts are shown in Figure~\ref{fig:specfit_un}.
While the reduced $\chi^2$ value of the fit to the photometry alone is typically better than that of the fit to the combined photometry and spectroscopy, with $\Delta \chi^2=0.3$, this is partially due to the inclusion of the spectroscopic data points.
The best fit models to the photometry alone do not match the observed spectra.

As the \zspec\ values are more than $2\sigma$ discrepant from the \zphot\ predictions, other derived properties for these galaxies can also be significantly different from the photometric predictions.
While the masses of all \sumg\ candidates with spectroscopic redshifts decrease from the photometric estimate, the galaxies are certainly still very massive, with all but the $z\sim1.2$ source having \logM~$>10.7$.
Of note, these galaxies all have considerable amounts of dust, $A_{\rm V}\sim3$~mag.
The derived properties from these fits are given in Table~\ref{tab:photspecprop} and Figure~\ref{fig:propcomp}.

Between their large amounts of dust and strong emission lines, the $z\sim2$ objects are likely dusty star-forming galaxies and perhaps sub-millimeter galaxies, which photometric redshift codes can have difficulty accurately identifying \citep[\eg][]{Caputi2012, Jin2024}.
Indeed one confirmed object, COS-DR1-209435, has a very strong 870 $\mu$m detection \citep[]{Simpson2020}.
When far-infrared data is available, fitting with photometric redshift and SED modeling programs which include such information can lead to better constraints on galaxy properties \citep[][Chang et al., in prep]{DaCunha2015, Dudzeviciute2020}.
Some of these objects may also be consistent with the hot dust-obscured galaxy (DOG) population \citep{Eisenhardt2012, Wu2012, Finnerty2020}.
Objects in that sample are identified from WISE colors, and often do not have extensive photometric observations similar to those in this work, which prevents a rigorous comparison of other parameters such as stellar mass or star formation rate.

The broad lines seen in five of the spectra, including all three confirmed at $z>3$, indicate AGN activity.
While the removal of this emission line flux from the photometry has reduced the stellar mass estimate, these galaxies all have stellar masses \logM~$>11.5$, with CDFS-VID1-2420 having an incredible stellar mass of \logM~$\sim11.9$.
However, while the emission line flux has been removed, the potential exists for significant AGN continuum flux contribution to the photometry, which could artificially inflate the stellar mass \citep[\eg][]{Furtak2024}.
The galaxies whose spectra show broad lines do appear compact in ground-based imaging, though several other objects with confirmed redshifts and no evidence for broad lines have similar morphologies, consistent with small sizes.

In an effort to account for this effect, we fit the combined photometric and spectroscopic data using \Bagpipes\ \citep{Carnall2018, Carnall2019b}, which includes functionality for fitting the AGN continuum contribution.
We also use a double power-law star-formation history, a \citet{Calzetti2000} dust law, and nebular emission lines, and allow corrections for spectral calibration and noise underestimation issues. 
The fits from \Bagpipes\ are in some cases marginally better fits to the data compared to those from FAST++, but they retain the high stellar masses and dust attenuation levels, despite estimating the contribution of AGN continuum and emission line flux.
As such we have confidence that these galaxies are indeed impressively massive, dust-obscured galaxies, though at lower redshifts than initially predicted from the photometry alone.

%------------------------------------------
%------------------------------------------

%------------------------------------------
\begin{figure*}
	\includegraphics[width=\textwidth, trim=0in 4.5in 1in 0in]{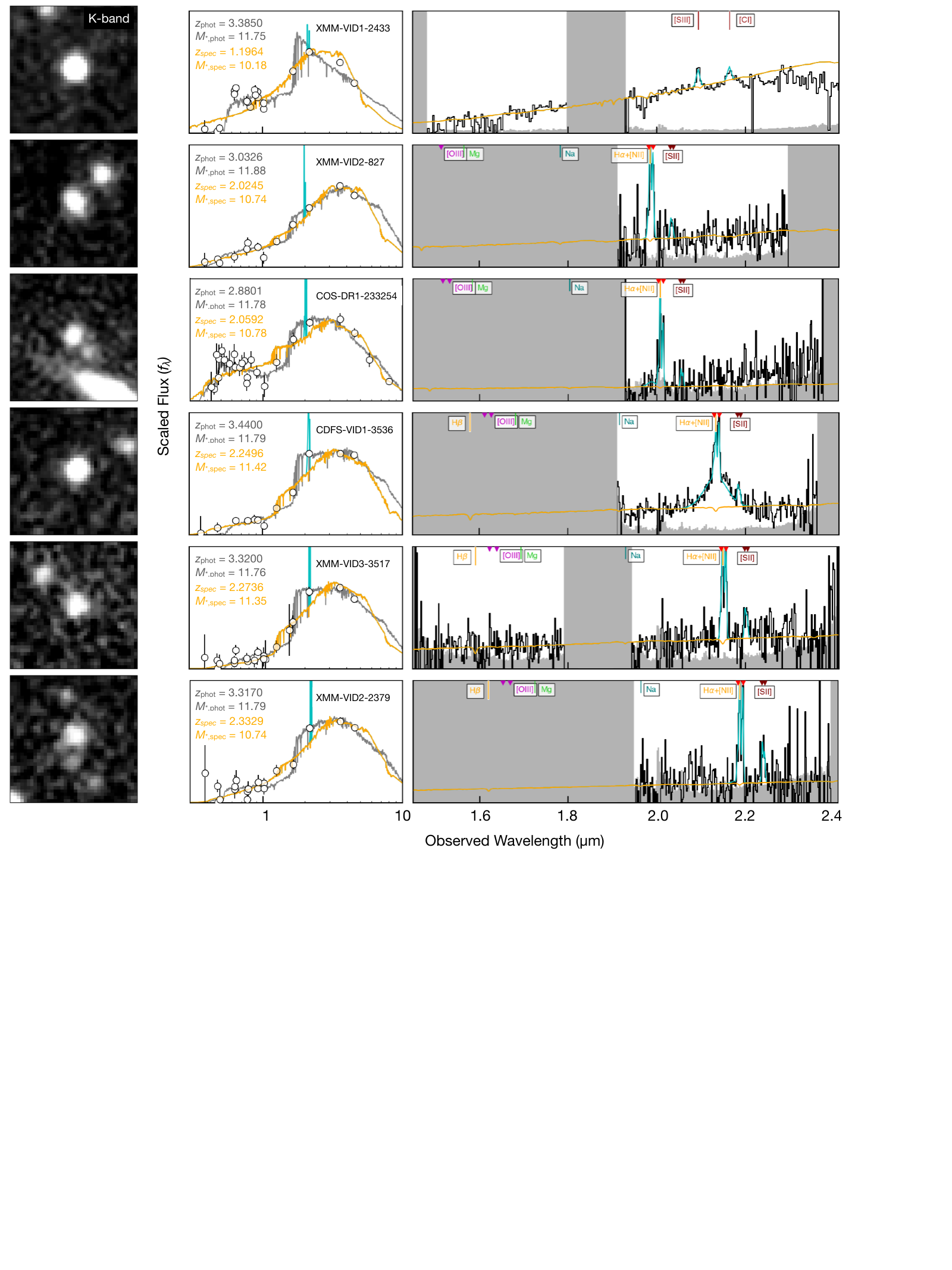}
    \caption{SUMG candidates with confirmed spectroscopic redshifts, in order of increasing \zspec.
    \textbf{Left:} $K$-band imaging cutouts of the target, 7.5\arcsec\ on a side.
    \textbf{Center:} The observed photometry of the galaxy (open circles) with best fit models to the photometry alone (gray line) and photometric and spectroscopic data (gold line with cyan emission lines).
    \textbf{Right:}  The observed spectrum binned in $\sim19.5\textrm{\AA}$ bins (12 pixels in $H$-band, 9 pixels in $K$-band) is shown in black, with uncertainty in gray. Completely gray regions were not observed. The emission line model is again cyan added to the best-fit continuum model in orange. Note that the vertical range in the center and right panels is not the same.}
    \label{fig:specfit1}
\end{figure*}
%------------------------------------------

\setcounter{figure}{2}

%------------------------------------------
\begin{figure*}
	\includegraphics[width=\textwidth, trim=0in 4.5in 1in 0in]{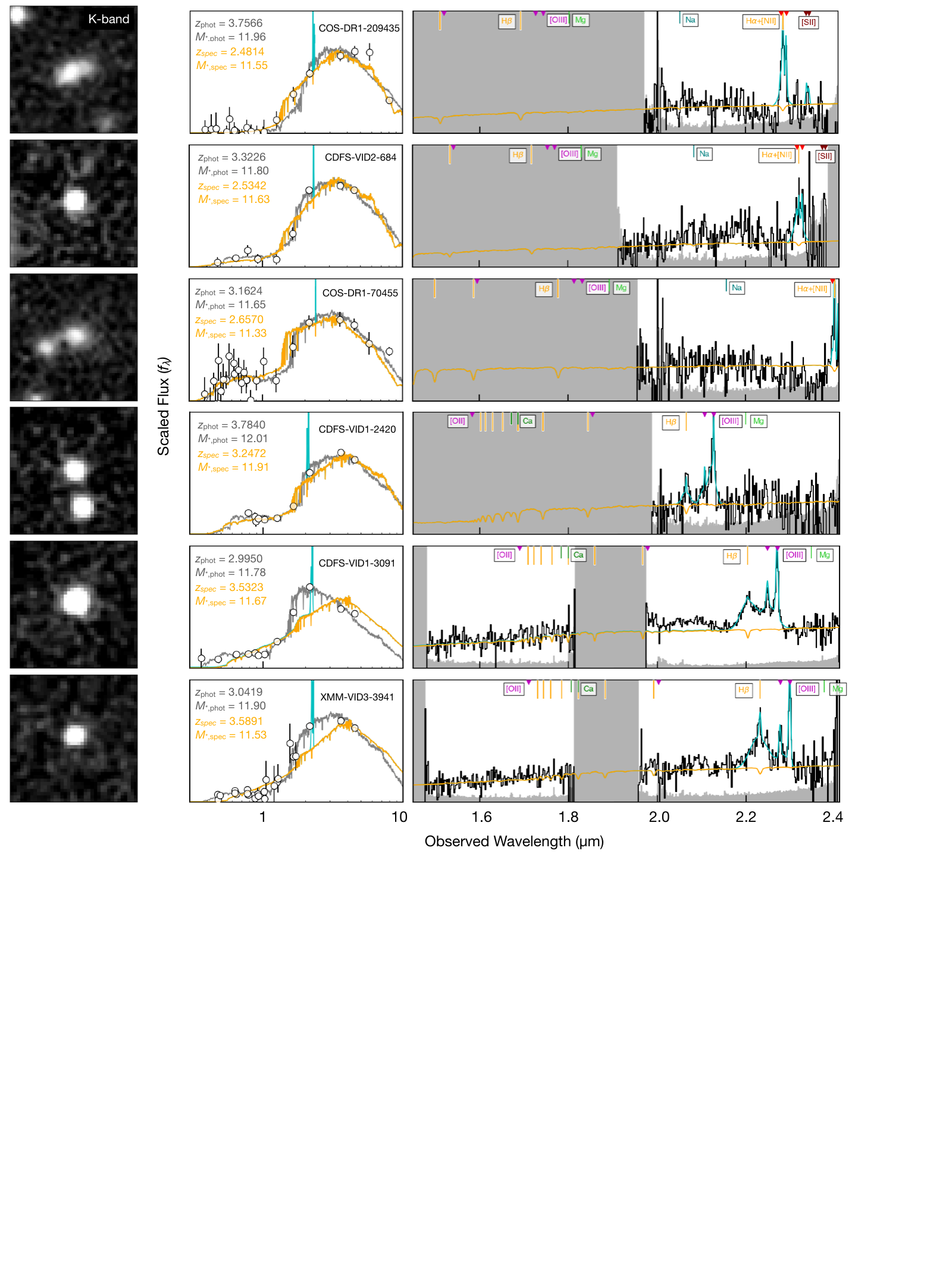}
    \caption{Continued.}
    \label{fig:specfit2}
\end{figure*}
%------------------------------------------

%------------------------------------------
\begin{figure*}
	\includegraphics[width=\textwidth, trim=0in 7.5in 1in 0in]{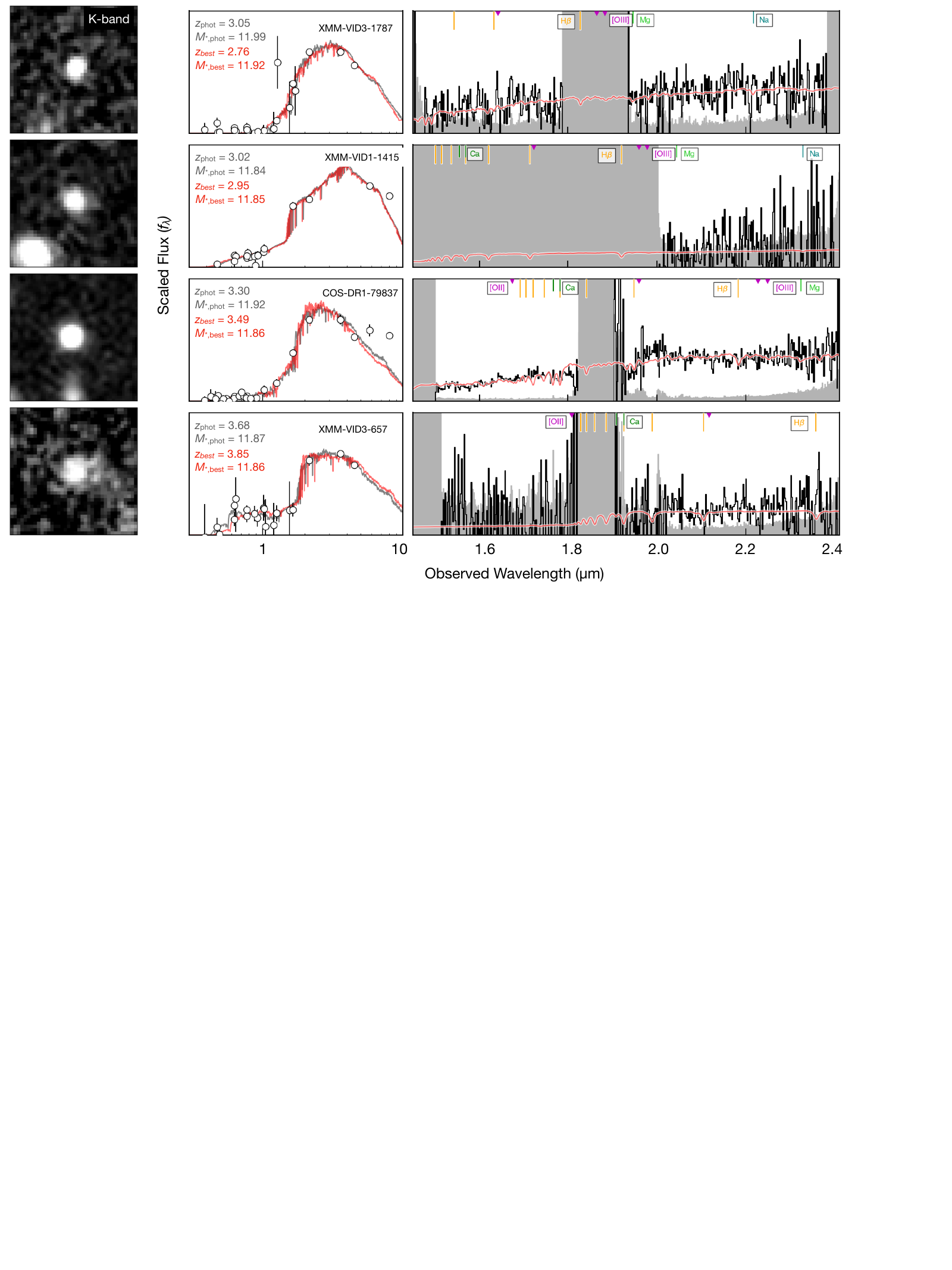}
    \caption{Same as Figure~\ref{fig:specfit2} for objects without confirmed spectroscopic redshifts. Best fit models to the combined photometry and spectroscopy are shown in red.}
    \label{fig:specfit_un}
\end{figure*}
%------------------------------------------

%------------------------------------------
\begin{table*}
	\centering
	\caption{Properties of the 16 spectroscopically-targeted candidate \sumg s. The first row for each object reports values derived from photometry alone (redshift, stellar mass, star formation rate, dust attenuation, lookback time to when 50\% of the stellar mass had formed, and lookback time to when the galaxy had quenched - SF indicates not yet quenched), while the second row shows the same variables recalculated using the spectroscopic redshift and line-flux corrected photometry. Galaxies marked with an asterisk are those showing broad lines from AGN activity.}
	\label{tab:photspecprop}
	\begin{tabular}{llcccccc}
		\hline
		Galaxy	& \zphot	& $M_{*,\rm phot}$	& SFR$_{\rm phot}$	& A$_{\rm V, phot}$	& $t_{\rm 50, phot}$	& $t_{\rm q, phot}$ \\
				& \zspec	& $M_{*,\rm spec}$	& SFR$_{\rm spec}$	& A$_{\rm V, spec}$ 	& $t_{\rm 50, spec}$	& $t_{\rm q, spec}$ \\
		 		&   		 &  log(M$_\odot$) 	&  log(M$_\odot$/yr)	&  mag			&  log(yr)			&  log(yr) \\
		\hline
		XMM-VID1-1415	& $3.02^{+0.09}_{-0.10}$	& $11.84^{+0.11}_{-0.08}$	& $2.56^{+0.29}_{-1.34}$	& $2.8^{+0.2}_{-0.1}$	& $8.13^{+0.33}_{-0.17}$	& $7.69^{+0.20}_{-7.69}$ \\
					& \hspace{2mm}-- & \hspace{2mm}--	& \hspace{2mm}-- & \hspace{2mm}-- & \hspace{2mm}-- & \hspace{2mm}-- \\
		XMM-VID1-2433	& $3.38^{+0.04}_{-0.04}$	& $11.75^{+0.03}_{-0.05}$	& $1.52^{+0.28}_{-0.25}$ 	& $1.3^{+0.2}_{-0.2}$	& $8.36^{+0.04}_{-0.14}$	& $8.08^{+0.01}_{-0.09}$ \\
					&  \hspace{2mm}1.1964	&  \hspace{2mm}$10.18^{+0.05}_{-0.02}$ 	&  \hspace{2mm}$2.46^{+0.04}_{-0.08}$ 	&  \hspace{2mm}$3.4^{+0.1}_{-0.0}$ 	&  \hspace{2mm}$7.24^{+0.08}_{-0.02}$ 	&  \hspace{2mm}SF\\		
		XMM-VID2-0827	& $3.03^{+0.22}_{-0.21}$	& $11.88^{+0.10}_{-0.58}$ & $2.98^{+0.18}_{-0.52}$ & $2.5^{+0.7}_{-0.2}$	& $8.81^{+0.08}_{-1.11}$	& \hspace{-1mm}SF \\
					& \hspace{2mm}2.0245 	& \hspace{2mm}$10.74^{+0.13}_{-0.06}$ 	& \hspace{2mm}$2.89^{+0.51}_{-0.24}$ 	& \hspace{2mm}$3.4^{+0.3}_{-0.1}$ 	& \hspace{2mm}$7.32^{+0.20}_{-0.29}$ 	& \hspace{2mm}SF \\
		XMM-VID2-2379	& $3.32^{+0.25}_{-0.25}$	& $11.79^{+0.01}_{-0.39}$	 & $2.01^{+2.57}_{-0.35}$	& $1.7^{+0.7}_{-0.2}$	& $8.90^{+0.16}_{-2.31}$	& $8.01^{+0.36}_{-8.01}$ \\
					&  \hspace{2mm}2.3329 	&  \hspace{2mm}$10.74^{+0.68}_{-0.04}$ 	&  \hspace{2mm}$3.74^{+0.13}_{-1.61}$ 	&  \hspace{2mm}$4.0^{+0.1}_{-1.2}$ 	&  \hspace{2mm}$6.73^{+2.19}_{-0.05}$ 	&  \hspace{2mm}SF \\
		XMM-VID3-0657	& $3.68^{+0.44}_{-0.32}$	& $11.87^{+0.04}_{-0.10}$	 & $1.76^{+0.08}_{-0.14}$ & $1.1^{+0.6}_{-0.3}$	& $9.07^{+0.03}_{-0.11}$	& $8.47^{+0.12}_{-0.36}$ \\
					& \hspace{2mm}--	& \hspace{2mm}--	& \hspace{2mm}--	& \hspace{2mm}-- & \hspace{2mm}--	& \hspace{2mm}-- \\
		XMM-VID3-1787	& $3.05^{+0.07}_{-0.39}$	& $11.99^{+0.00}_{-0.11}$	 & $<0$ 				& $1.0^{+0.4}_{-0.1}$ 	& $9.25^{+0.01}_{-0.06}$	& $9.18^{+0.06}_{-0.02}$ \\
					& \hspace{2mm}-- & \hspace{2mm}-- & \hspace{2mm}-- & \hspace{2mm}-- & \hspace{2mm}-- & \hspace{2mm}-- \\
		XMM-VID3-3517	& $3.27^{+0.18}_{-0.19}$	& $11.76^{+0.05}_{-0.10}$	 & $<0$ 				& $2.0^{+0.5}_{-0.2}$	& $8.39^{+0.07}_{-0.24}$ 	& $8.25^{+0.05}_{-0.27}$ \\
					&  \hspace{2mm}2.2736 	&  \hspace{2mm}$11.35^{+0.02}_{-0.04}$ 	&  \hspace{2mm}$0.70^{+0.78}_{-0.07}$ 	&  \hspace{2mm}$3.3^{+0.1}_{-0.0}$ 	&  \hspace{2mm}$7.98^{+0.04}_{-0.06}$ 	&  \hspace{2mm}$7.78^{+0.01}_{-0.10}$ \\
		XMM-VID3-3941*	& $3.04^{+0.12}_{-0.11}$	& $11.90^{+0.01}_{-0.10}$ & $1.50^{+0.04}_{-0.31}$ & $1.4^{+0.0}_{-0.3}$	& $9.13^{+0.00}_{-0.10}$ 	& $8.69^{+0.06}_{-0.10}$ \\
					&  \hspace{2mm}3.5891 	&  \hspace{3mm}$11.53^{+0.02}_{-0.01}$	&  \hspace{2mm}$4.75^{+0.08}_{-0.03}$ 	&  \hspace{2mm}$4.0^{+0.0}_{-0.0}$ 	&  \hspace{2mm}$6.53^{+0.02}_{-0.04}$ 	&  \hspace{2mm}SF \\
	\hline
		CDFS-VID1-2420*	& $3.78^{+0.21}_{-0.27}$ & $12.01^{+0.15}_{-0.05}$	& $2.14^{+0.70}_{-0.24}$	& $1.5^{+0.8}_{-0.2}$ 	& $9.00^{+0.07}_{-0.13}$	& $8.11^{+0.36}_{-8.11}$ \\
					&  \hspace{2mm}3.2472 	&  \hspace{3mm}$11.91^{+0.00}_{-0.05}$ 	&  \hspace{2mm}$2.66^{+0.22}_{-0.18}$ 	&  \hspace{2mm}$2.3^{+0.2}_{-0.2}$ 	&  \hspace{2mm}$8.97^{+0.05}_{-0.11}$ 	&  \hspace{2mm}SF \\			
		CDFS-VID1-3091*	& $3.00^{+0.17}_{-0.21}$	& $11.78^{+0.06}_{-0.04}$ & $1.26^{+0.19}_{-0.37}$	& $0.8^{+0.2}_{-0.4}$	& $9.00^{+0.13}_{-0.03}$	& $8.65^{+0.13}_{-0.06}$ \\
					&  \hspace{2mm}3.5323 	&  \hspace{3mm}$11.67^{+0.00}_{-0.00}$ 	&  \hspace{2mm}$4.99^{+0.00}_{-0.02}$ 	&  \hspace{2mm}$3.8^{+0.0}_{-0.0}$ 	&  \hspace{2mm}$6.47^{+0.01}_{-0.00}$ 	&  \hspace{2mm}SF \\			
		CDFS-VID1-3536*	& $3.44^{+0.19}_{-0.40}$	& $11.79^{+0.08}_{-0.24}$ & $0.43^{+2.05}_{-1.31}$	& $2.2^{+1.7}_{-0.2}$	& $8.24^{+0.68}_{-1.76}$	& $8.07^{+0.62}_{-8.07}$ \\
					&  \hspace{2mm}2.2496 	&  \hspace{3mm}$11.42^{+0.04}_{-0.08}$ 	&  \hspace{2mm}$2.53^{+0.18}_{-0.20}$ 	&  \hspace{2mm}$2.9^{+0.2}_{-0.2}$ 	&  \hspace{2mm}$8.81^{+0.12}_{-0.26}$ 	&  \hspace{2mm}SF \\
		CDFS-VID2-0684*	& $3.32^{+0.28}_{-0.28}$	& $11.80^{+0.11}_{-0.08}$ & $1.51^{+0.17}_{-0.10}$	& $1.4^{+0.6}_{-0.4}$ 	& $9.03^{+0.10}_{-0.50}$	& $8.59^{+0.16}_{-0.52}$ \\
					&  \hspace{2mm}2.5342 	&  \hspace{3mm}$11.63^{+0.03}_{-0.03}$ 	&  \hspace{2mm}$1.89^{+0.26}_{-0.21}$ 	&  \hspace{2mm}$2.4^{+0.2}_{-0.2}$ 	&  \hspace{2mm}$9.03^{+0.07}_{-0.10}$ 	&  \hspace{2mm}SF \\
		\hline
		COS-DR1-70455	& $3.15^{+0.26}_{-0.23}$	& $11.75^{+0.03}_{-0.16}$ & $1.91^{+0.26}_{-0.15}$	& $1.5^{0.5}_{-0.3}$		& $9.03^{+0.07}_{-0.23}$	& \hspace{-1mm}SF \\
					&  \hspace{2mm}2.6570 	&  \hspace{2mm}$11.33^{+0.16}_{-0.16}$ 	&  \hspace{2mm}$2.28^{+0.30}_{-0.33}$ 	&  \hspace{2mm}$2.2^{+0.3}_{-0.4}$ 	&  \hspace{2mm}$8.62^{+0.50}_{-0.49}$ 	&  \hspace{2mm}SF \\
		COS-DR1-79837	& $3.32^{+0.17}_{-0.22}$	& $11.92^{+0.06}_{-0.02}$ & $0.63^{+0.65}_{-0.03}$	& $0.8^{+0.7}_{-0.2}$	  	& $9.12^{+0.00}_{-0.30}$	& $8.89^{+0.00}_{-0.21}$ \\
					& 	 \hspace{2mm}--	& 	 \hspace{2mm}--	&  \hspace{2mm}--		&  \hspace{2mm}--		&  \hspace{2mm}--		&  \hspace{2mm}-- \\
		COS-DR1-209435	& $3.77^{+0.33}_{-0.37}$	& $11.98^{+0.02}_{-0.20}$ & $<0$ 				& $0.9^{+0.4}_{-0.2}$ 	& $9.20^{+0.05}_{-0.21}$	& $9.19^{+0.05}_{-0.20}$ \\
					&  \hspace{2mm}2.4814 &  \hspace{2mm}$11.55^{+0.04}_{-0.09}$ &  \hspace{2mm}$<1.1$ & \hspace{2mm}$2.9^{+0.1}_{-0.3}$ &  \hspace{2mm}$8.63^{+0.18}_{-0.17}$ &  \hspace{2mm}$8.58^{+0.18}_{-0.30}$\\
		COS-DR1-233254	& $2.88^{+0.17}_{-0.10}$	& $11.81^{+0.03}_{-0.12}$ & $2.08^{+0.28}_{-0.13}$	& $1.3^{+0.5}_{-0.3}$	& $9.13^{+0.00}_{-0.13}$	& \hspace{-1mm}SF \\
					&  \hspace{2mm}2.0592	&  \hspace{2mm}$10.78^{+0.08}_{-0.03}$ 	&  \hspace{2mm}$2.42^{+0.55}_{-0.19}$ 	&  \hspace{2mm}$2.6^{+0.2}_{-0.1}$ 	&   \hspace{2mm}$7.53^{+0.14}_{-0.07}$	&  \hspace{2mm}SF \\
	\end{tabular}
\end{table*}
%------------------------------------------

%------------------------------------------
\begin{figure*}
	\includegraphics[width=\textwidth, trim=0in 2in 4in 0in]{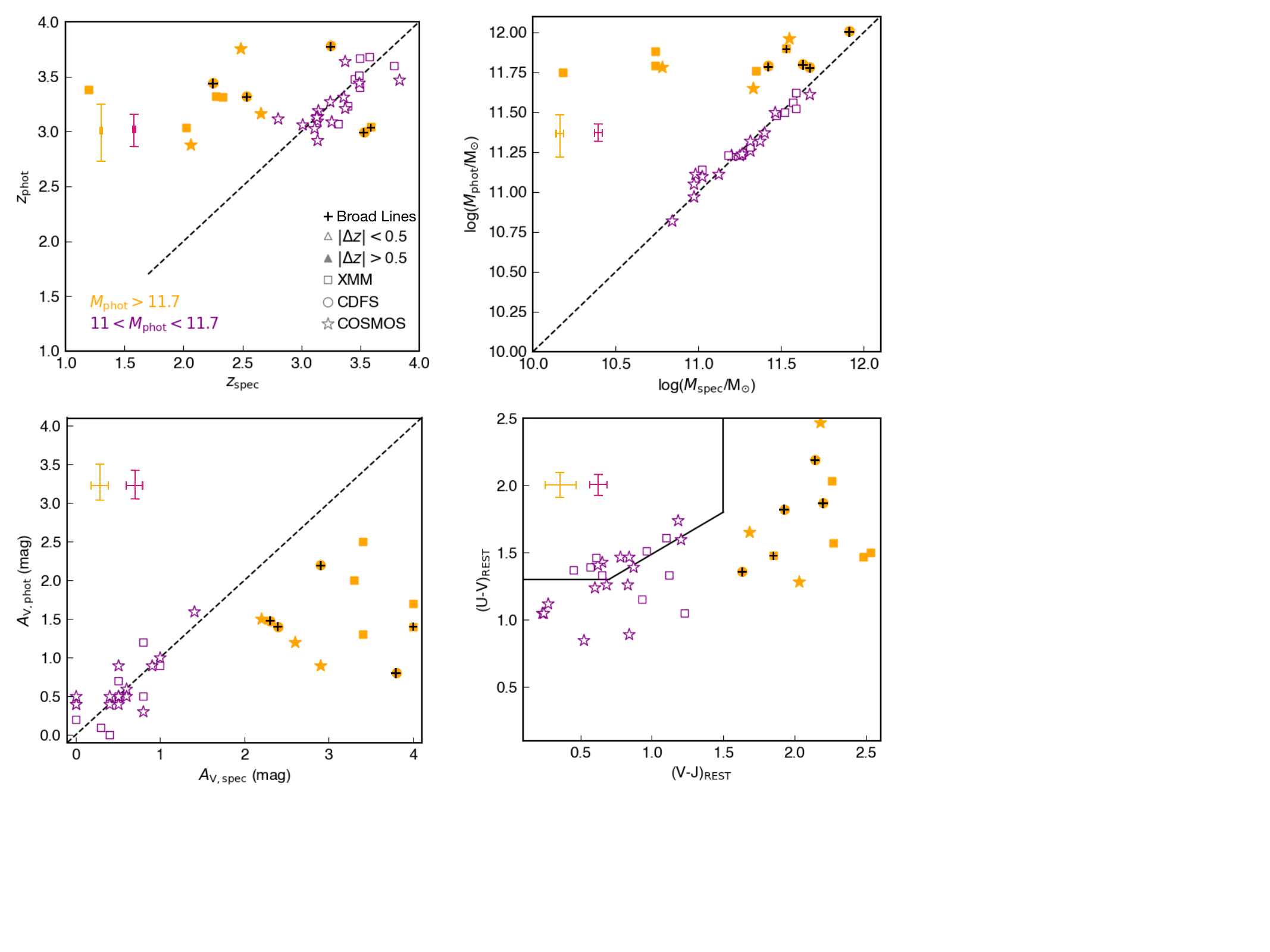}
    \caption{Comparison of galaxy properties with and without spectroscopic information - redshifts (top left), stellar masses (top right), and dust attenuation values (bottom left) are shown.
    Marker shape indicates the field, color indicates the stellar mass estimated from photometry alone, and a filled symbol represents objects with $\lvert$~\zphot-\zspec~$\rvert>0.5$. A clear discrepancy is seen between the UMG candidates (studied in previous works; purple points), which have generally good agreement, and the \sumg\ candidates (the focus of this paper; yellow markers), which typically have different redshifts, lower stellar masses, and greater dust attenuation than predicted from photometry. Spectra exhibiting broad lines from AGN activity are marked with a cross.
    The bottom right panel shows the spectroscopically derived rest-frame colors for comparison with Figure~\ref{fig:samp}. The redshift interlopers have extremely red rest-frame colors.
    }
    \label{fig:propcomp}
\end{figure*}
%------------------------------------------

\clearpage

\section{Comparison of Photometric Candidate Properties in COSMOS}\label{Sec:AnComp}

Modeling of galaxy SEDs to obtain stellar masses requires both sufficient photometric sampling of the SED and flexible, accurate models to fit the observations.
Extragalactic fields such as COSMOS, \mbox{GOODS-S} (CDFS), and UDS (XMM), among others, have sufficient multi-wavelength observations from ultraviolet to near-infrared wavelengths to constrain redshifts and stellar masses well for most galaxies, though more bandpasses, and in particular medium- and narrow-bandpasses can improve the uncertainties on these measurements \citep[\eg][]{Whitaker2011, Sobral2013, Suzuki2015, Straatman2016, Sarrouh2024}.
Also critical to accurate photometric redshifts and stellar masses is the modeling of strong emission lines from star formation and/or AGN \citep[\eg][]{Stark2013, Smit2014, Salmon2015, Forrest2017, Withers2023}.

Particularly at higher redshifts, near-infrared photometry is required to probe redward of the \Dfour\ feature and constrain the mass.
Deep IRAC data has historically been necessary to avoid the potential of a single line in the $K$-band leading to overestimates in stellar mass.
While the improved sensitivity and spatial resolution of JWST may eventually preclude the need for Spitzer data to constrain the near-infrared in some fields, JWST's smaller field of view means that deep IRAC data will still be a critical part of large area catalogs for the foreseeable future.

The catalogs used for target selection in this work satisfy these requirements and show excellent agreement between photometric and spectroscopic redshifts for galaxy populations as a whole.
Additionally, they have reported completeness limits ($\geq 80\%$) down to at least 1 magnitude fainter in $K_s$ than the faintest target observed in this program, yet a significant fraction of the \sumg\ candidates do not have redshifts and masses consistent with the photometric estimates.
In what follows we attempt to uncover the root of this issue by comparing objects in the COSMOS field, for which numerous deep photometric catalogs have been constructed.

\subsection{Catalogs Considered}

The stellar masses of the galaxies in question determined from modeling the photometry alone are exceptionally high.
In order to confirm that this is not a result of the particular methods used in the generation of the UltraVISTA catalogs, we compare results between the UltraVISTA DR1 \citep{Muzzin2013a}, UltraVISTA DR3 (A. Muzzin, private communication), COSMOS2015 \citep{Laigle2016}, and both sets of COSMOS2020 \citep[CLASSIC and FARMER;][]{Weaver2022} photometric catalogs (5 in total).
This allows for analysis of differences in galaxy characterization due to different imaging bandpasses and depths (different publications), different methods of photometric extraction using the same images (COSMOS2020 CLASSIC vs. FARMER), and different SED modeling codes on the same photometry (COSMOS2020 eazy-py and LePhare characterizations).

An in depth discussion of the differences of the methodologies involved in the construction of these catalogs is beyond the scope of this work, and we refer the interested reader to the relevant publications for more details.
Briefly however, the UltraVISTA DR1 and UltraVISTA DR3 catalogs use \texttt{SExtractor} \citep{Bertin1996} on deep $K$-band images to identify sources and measure photometry and MOPHONGO \citep{Labbe2006, Labbe2010} to deblend IRAC photometry based on the $K$-band image, assuming no color gradients from the $K$-band to the Spitzer bandpasses.
The COSMOS2015 and COSMOS2020 CLASSIC catalogs use \texttt{SExtractor} on a combined $zY JHK_s$ and $izYJHK_s$ detection image, respectively, for the source detection and IRACLEAN \citep{Hsieh2012} to deblend IRAC photometry.
The COSMOS2020 FARMER catalog uses \texttt{The Tractor} \citep{Lang2016} to extract photometry from images in different bandpasses simultaneously.
Differences in detection image bandpass and depth, as well as the specifics of source detection inputs, can influence not only whether a source is detected or not, but also the total integrated magnitude of the source in each bandpass.

Photometric redshifts for the UltraVISTA DR1 and UltraVISTA DR3 catalogs are derived using EAzY \citep{Brammer2008} and stellar population parameters are derived using FAST \citep{Kriek2009}.
One set of photometric redshifts in both the COSMOS2020 CLASSIC and FARMER catalogs are derived using the similar eazy-py code \citep{Brammer2021}, which also models stellar population parameters.
COSMOS2015 and a second set of photometric redshifts and stellar population parameters in both COSMOS2020 catalogs are derived using LePhare \citep{Arnouts1999, Ilbert2006}, incorporating the BC03 models \citep{Bruzual2003}.
As a result, for an object detected in all 5 catalogs, there are 7 characterizations of its redshift and stellar populations.

%------------------------------------------
\begin{table*}
	\centering
	\caption{Comparison of photometric properties for four targeted \sumg s in the COSMOS field. Catalogs considered are the UltraVISTA DR1 \citep[UVDR1;][]{Muzzin2013}, COSMOS2015 \citep[C15;][]{Laigle2016}, COSMOS2020 CLASSIC (C20C) and COSMOS2020 FARMER \citep[C20F;][]{Weaver2022}. None of these objects are in the ultra-deep stripes of the proprietary UltraVISTA DR3 catalog.}
	\label{tab:obscomp}
	\begin{tabular}{llrrrr}
		\hline
		Property & Source &   &  &  &  \\
		\hline
		\hline
		ID		& UVDR1			& 70455 	& 79837 	& 209435 	& 233254 \\
				& C15			& 440292	& 530898	& 798016	& 985682	\\
				& C20C			& 567661	& 718745	& 1152024 & 1450899 \\
				& C20F			& --		& 768111	& 670662	&  --	\\
		\hline
	$\Delta_{\rm proj, UVDR1}$ (\arcsec) & C15  & 0.10  & 0.02  & 0.08  & 0.31  \\
				& C20C			& 0.18	& 0.14	& 0.12 	& 0.32  \\
				& C20F			& --		& 0.13	& 0.05	&  --	\\
		\hline
		$m_K$	& UVDR1			& 21.85  	& 21.10	& 22.03	&  21.10	\\
				& C15			& 22.13	& 21.05	& 21.99	&  21.76	\\
				& C20C			& 21.94	& 21.10	& 21.97	&  20.14	\\
				& C20F			&  --		& 21.03	& 21.92	&  --  	\\		
		\hline
		$z$	 	& UVDR1 EAZY 	& $3.149^{+0.259}_{-0.230}$	& $3.318^{+0.168}_{-0.214}$	& $3.769^{+0.332}_{-0.365}$	&  $2.870^{+0.177}_{-0.152}$	\\
				& C15	LePhare	& $2.924^{+0.379}_{-0.240}$	& $2.783^{+0.118}_{-0.114}$	& $3.259^{+0.314}_{-0.139}$	&  $2.773^{+0.295}_{-0.642}$	\\
				& C20C	eazy-py 	& $3.401^{+0.064}_{-0.070}$	& $2.452^{+0.387}_{-0.066}$	& $2.439^{+0.207}_{-0.240}$	&  $11.83^{+0.076}_{-0.116}$	\\
		  		& C20C	LePhare 	& $2.639^{+0.171}_{-0.178}$	& $2.883^{+0.123}_{-0.138}$	& $3.460^{+0.232}_{-0.298}$	&  $2.783^{+0.092}_{-0.302}$	\\
		    		& C20F	eazy-py	&  --		& $1.238^{+0.014}_{-0.029}$	& $0.050^{+0.084}_{-0.060}$	&  --	\\
		      		& C20F	LePhare 	&  --		& $1.201^{+0.106}_{-0.136}$	& $0.722^{+0.018}_{-0.040}$	&  --	\\
		& \textbf{Spec FAST++} &	\textbf{2.6570} & \textbf{--	} & \textbf{2.4814} & \textbf{2.0592} \\		
		\hline
		log($M$/M$_\odot$)	& UVDR1 FAST & 11.75  & 11.99  & 11.98  & 11.81	\\
				& C15	LePhare	& 11.32	& 12.32	& 11.92	& 11.60	\\
				& C20C	eazy-py 	& 11.41	& 11.57	& 11.75	& nan	\\
		  		& C20C	LePhare 	& 11.26	& 12.02	& 11.95	& 11.47	\\
		    		& C20F	eazy-py	&  --		& 10.09	& 5.94	&  --	\\
		      		& C20F	LePhare 	&  --		& 9.75	& 7.99	&  --	\\
		& \textbf{Spec FAST++} &	\textbf{11.33}  & \textbf{--}	& \textbf{11.55}	& \textbf{10.78}	\\
		\hline
	   log(SFR/($M_\odot$/yr)) & UVDR1 FAST & 1.82	& 1.36  & $< 0$	& 2.08	\\
				& C15	LePhare	& 2.30	& 2.83	& 2.45	& 2.13	\\
				& C20C	eazy-py 	& 3.04	& 3.21	& 2.94	& nan	\\
		  		& C20C	LePhare 	& 1.96	& 2.43	& 2.55	& $< 0$	\\
		    		& C20F	eazy-py	&  --		& 0.12	& $< 0$	&  --	\\
		      		& C20F	LePhare 	&  --		& 0.72	& 0.18	&  --	\\
		& \textbf{Spec FAST++} &	\textbf{2.28}  & \textbf{--}  & \textbf{$<1.1$}  & \textbf{2.42} \\
		\hline
		$A_{\rm V}$	& UVDR1 FAST & 1.4	& 1.5		& 0.9		& 1.3	\\
				& C15	LePhare	& --		&  --		&  --		&  --	\\
				& C20C	eazy-py 	& 2.4		& 3.6		& 3.8		& nan \\
		  		& C20C	LePhare 	&  --		&  --		&  --		&  --	\\
		    		& C20F	eazy-py	&  --		& 0.6		& 1.5		&  --	\\
		      		& C20F	LePhare 	&  --		&  --		&  --		&  --	\\
		& \textbf{Spec FAST++} &	\textbf{2.2}  & \textbf{--}  & \textbf{2.9}  & \textbf{2.6}	\\		
		\hline
	\end{tabular}
\end{table*}
%------------------------------------------

\subsection{Observed \sumg\ Candidates}

In addition to many of the UMGs confirmed in \citet{Forrest2020b}, four of the \sumg\ candidates which were targeted spectroscopically are in the COSMOS field, and all were selected from the UltraVISTA DR1 catalog (they are covered by the ultra-deep stripes in the UltraVISTA DR3 catalog).
We search for the counterparts of these objects in the COSMOS2015 and COSMOS2020 catalogs using a projected radius of $0.5\arcsec$ and, in the case of multiple such matches, select the neighbor within that radius with the most similar $K$-band magnitude.
The four \sumg\ candidates are detected in the COSMOS2015 and COSMOS2020 CLASSIC catalogs, while only two of them (COS-DR1-79837 and COS-DR1-209435) are detected in the COSMOS2020 FARMER catalog (COS-DR1-70455 and COS-DR1-233254 are not).
The associated derived properties for these objects in the various catalogs are given in Table~\ref{tab:obscomp}.

The FARMER catalog does not detect COS-DR1-70455 and COS-DR1-233254 due to their proximity to bright stars, the areas around which are masked more conservatively than in other catalogs.
It is notable that these two objects show considerably more variance in their reported $K_s$-band magnitudes between catalogs.
While COS-DR1-79837 and COS-DR1-209435 have a range of $\Delta m_K\lesssim0.1$ mag between catalogs, COS-DR1-70455 and COS-DR1-233254 have ranges of $\Delta m_K=0.3$ and 1.6 mag, respectively.
This suggests that differences in accounting for flux from the nearby stars is contributing to variance in the extracted magnitudes, though in the case of the COS-DR1-233254 several somewhat close neighbors are visible in imaging which may also have been subsumed into the flux totals for the counterpart of this object in some catalogs - reproducing and analyzing this aspect of source detection is beyond the scope of this work.
Both characterizations of COS-DR1-79837 and COS-DR1-209435 in the FARMER catalog yield significantly lower redshifts (and therefore lower stellar mass estimates) than the other parameterizations, and in the case of COS-DR1-209435, considerably lower than the \zspec.

While some off these objects have redshifts and stellar population parameters derived from the photometry which are similar to those derived once the redshift is fixed to the spectroscopic redshift, these values are more often discrepant, and no catalog produces consistently better agreement than the others.
Given that in general the photometry for these objects is roughly consistent between catalogs, this suggests that the main difficulty with these objects is obtaining an accurate photometric redshift, which then propagates into determinations of stellar mass, etc.
Of course, a small sample size is considered here, and photometric redshift codes may not be optimized for describing rare objects accurately when generating large catalogs.
Also, we note that these large discrepancies were not seen for the lower mass (and generally bluer) UMGs described in \citet{Forrest2020b}.

\subsection{Broader Comparison of (S)UMG Candidates between Catalogs }

This raises the question of not only how reliably, but also how consistently, high mass galaxy candidates at these redshifts can be identified photometrically.
To explore this, in each characterization we find all UMG (\logM~$>11.0$, $3<$~\zphot~$<4$) and \sumg\ candidates (\logM~$>11.7$, $3<$~\zphot~$<4$) with $m_K<23.4$, the brightest reported completeness limit of the catalogs (80\%).
Note that in these cases no further cuts are made for model-SED agreement, neighbor contamination, etc.

In Tables~\ref{tab:count_umg} \& \ref{tab:count_sumg} we show the number of identified UMG and \sumg\ candidates in each catalog.
We then match these objects to their sources in the other catalogs as above, finding neighbors with coordinates within $0.5\arcsec$ of those in the initial catalog and choosing the one which is most similar in $K$-band magnitude when there are multiple matches.
We report the number of these candidates which have a counterpart in at least one other catalog, and the number which have a counterpart in at least three other catalogs (out of the five considered).
This removes the requirement of a target lying in the UltraVISTA DR3 footprint, which covers roughly half the area of the other catalogs.
Similarly, we report the number of those counterparts which also have photometric redshifts and stellar masses satisfying either UMG or \sumg\ cuts in at least one other stellar population characterization, and the number of counterparts satisfying these cuts in at least five other characterizations (out of the seven considered).

Of all the UMG candidates, $96.9\%$ have counterparts in at least one other catalog and $53.0\%$ are characterized as UMGs in at least one other characterization.
For the more restrictive comparison, $84.5\%$ have counterparts in at least three other catalogs, but only $5.0\%$ are characterized as UMGs in at least five other characterizations.
Of all the \sumg\ candidates, $87.9\%$ have counterparts in at least one other catalog and $24.8\%$ are characterized as UMGs in at least one other characterization.
Similarly, $73.0\%$ have counterparts in at least three other catalogs, but \textit{none} are characterized as UMGs in at least five other characterizations.
COSMOS2020 tends to predict more high mass galaxies than other catalogs, but the COSMOS2020 Classic LePhare characterization yields lower numbers than UltraVISTA DR1 for instance.
The fraction of candidates in other catalogs is also higher for COSMOS2020 since both the CLASSIC and FARMER catalogs used the same detection image.

\begin{figure*}
    \includegraphics[width=\textwidth, trim=0 3.5in 0 0]{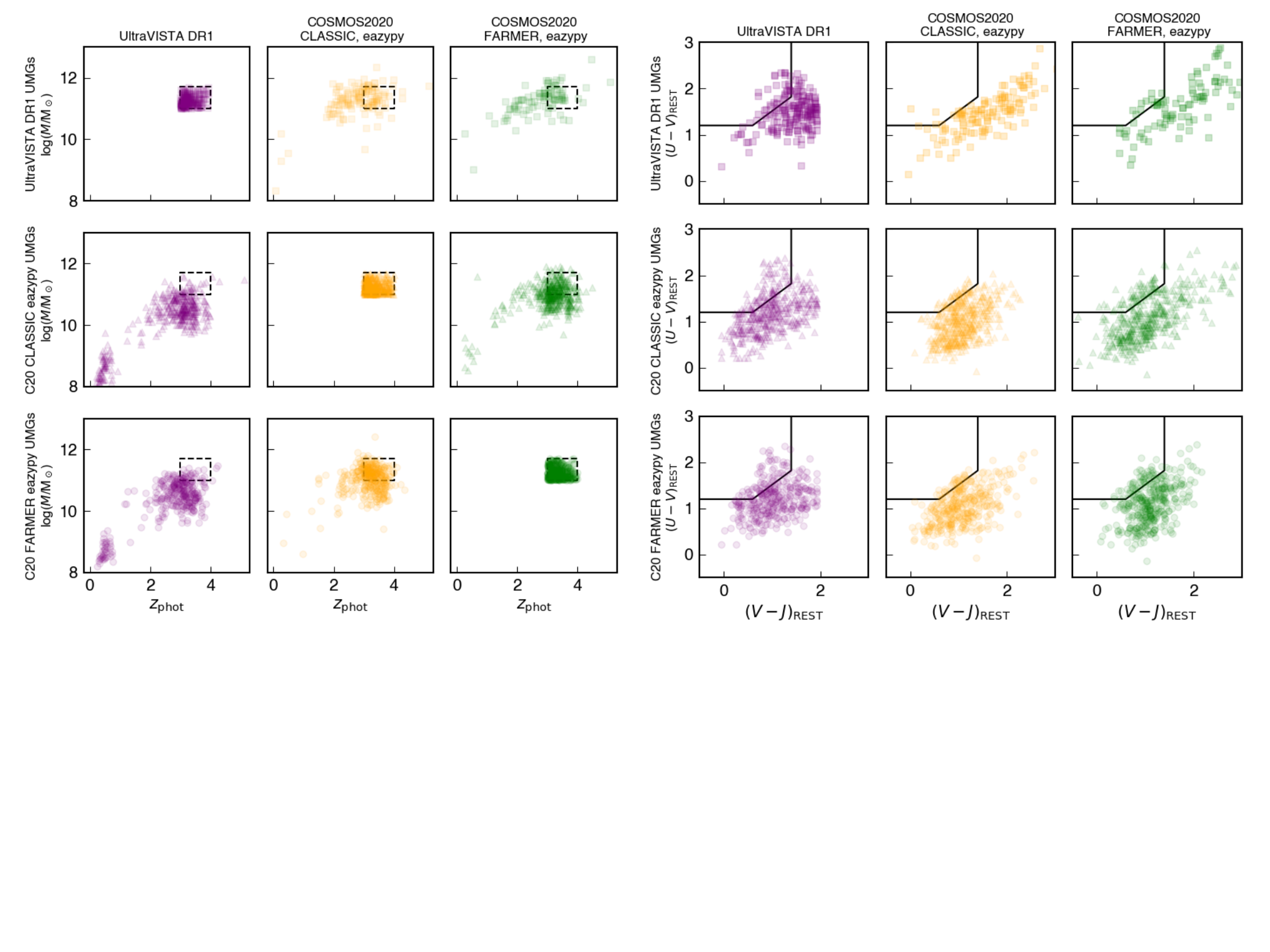}
    \caption{Comparison of UMGs selected from different photometric catalogs. \textbf{Left side:} The photometric redshifts and stellar masses of UMGs. The selection is represented by a black dashed box for comparison.
    In each row, either the UltraVISTA DR1 (top), COSMOS2020 CLASSIC eazypy (middle), or COSMOS2020 FARMER eazypy (bottom) catalog is used to select UMGs. The properties of the best matches to these objects in the other catalogs are shown in the other panels in that row.
    Objects with \zphot$<1$ are ostensibly due to photometric confusion between the Lyman- and Balmer-break features. Note that not all sources have a match in all catalogs.
     \textbf{Right side:} The location of UMGs on the restframe \UVJ\ color-color diagram following the same scheme as the left side.}
    \label{fig:UMGcomp}
\end{figure*}

\begin{figure*}
    \includegraphics[width=\textwidth, trim=0 3.5in 0 0]{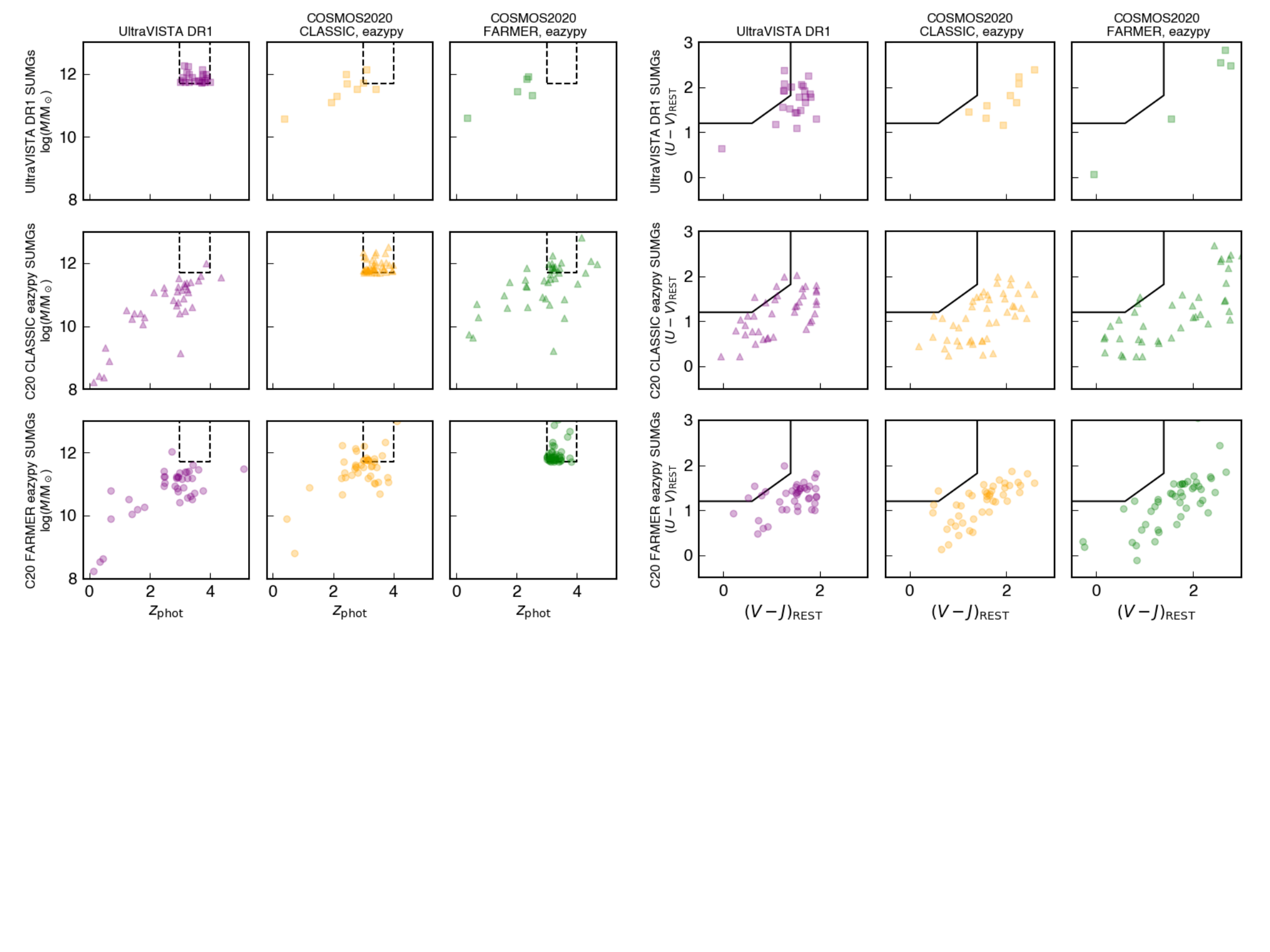}
    \caption{The same as Figure~\ref{fig:UMGcomp} but for the \sumg\ sample selection (\logM~$>11.7$).}
    \label{fig:SUMGcomp}
\end{figure*}

The photometric redshifts, stellar masses, and rest-frame colors of UMG and \sumg\ candidates selected from UltraVISTA DR1 and the COSMOS2020 eazypy characterizations from the CLASSIC and FARMER catalogs are shown in Figures~\ref{fig:UMGcomp} \& \ref{fig:SUMGcomp}, along with those of their counterparts in the other characterizations.
When UMG and \sumg\ candidates from one characterization are not classified as such in another characterization, they generally have slightly lower photometric redshifts ($2<z<3$) and stellar masses (\logM~$>10.5$), though the rest-frame UVJ colors are often quite red.
Notably, the UltraVISTA DR1 EAZY template set did not contain templates as red as the COSMOS2020 eazy-py template set, limiting derived rest-frame colors to \mbox{\VJc~$<2$}.
We note that for all of the \sumg\ candidates spectroscopically observed, even in the most extreme cases of effects that can confound photometric redshift codes such as high dust attenuation and strong emission line contamination, all objects but one with spectroscopic redshifts still have \logM~$>10.7$ and \zspec~$>2$.
There are a handful of $z>3$ photometric candidates which other characterizations place at $z<1$ as well, potentially due to the choice of interpretation of a photometric break as either a $z>3$ Lyman break or a $z<1$ Balmer break.

From this analysis there does not appear to be one catalog and SED modeling method 
which is definitively better than the others in terms of UMG candidate selection.
This suggests that while different methods of source detection, masking of regions around stars, etc. do play a role in differences between catalogs, different photometric redshift modeling codes and their associated fitting templates are the most substantial source of discrepancy in massive galaxy candidate identification.

%------------------------------------------
\begin{table*}
	\centering
	\caption{Comparison of candidate UMGs in the COSMOS field between different catalogs. Here we consider objects with $3<$~\zphot~$<4$, $11.0<$~\logM~$<11.7$, and $m_K<23.4$. There are 5 catalogs considered for counterpart detection (UltraVISTA DR1 and DR3, COSMOS2015, COSMOS2020 CLASSIC and FARMER) and 7 object characterizations (the eazy-py and Lephare characterizations for both COSMOS2020 catalogs). }
	\label{tab:count_umg}
	\begin{tabular}{lcrrrrr}
		\hline
		Catalog  & Characterization	& Total UMG  & Counterparts 	& Counterparts 		& Counterparts	which are	& Counterparts which are	 \\
			   &					& Candidates   & in 1+ Other   	& in 3+ Other   		& UMGs in 1+ Other		& UMGs in 5+ Other 	    \\
			   &					& 			  & Catalogs		& Catalogs		& Characterizations   	& Characterizations    \\
		\hline
		\hline
		DR1 	  & EAZY+FAST	& 201   & 155	& 108  &   85   &   14   \\
		DR3 	  &  EAZY+FAST 	&   97   &   91	&   76  &   70   &     9   \\
		C15 	  & LePhare		& 188   & 183	& 150  & 114   &   16 \\
		C20C &  eazy-py	    	& 476   & 476	& 422  & 220   &   12 \\
		C20C &  LePhare		& 148   & 148 	& 133  & 116   &   15 \\
		C20F &  eazy-py	     	& 375   & 375	& 338  & 202   &   11 \\
		C20F &  LePhare	     	& 369   & 369	& 339  & 176   &   15 \\
	\end{tabular}
\end{table*}
%------------------------------------------

%------------------------------------------
\begin{table*}
	\centering
	\caption{Comparison of massive galaxies in the COSMOS field between different catalogs. Here we consider objects with $3<$~\zphot~$<4$ and $m_K<23.4$, with UMGs having \logM~$\ge 11.0$ and \sumg s having \logM~$\ge 11.7$. There are 5 catalogs considered for counterpart detection (UltraVISTA DR1 and DR3, COSMOS2015, COSMOS2020 CLASSIC and FARMER) and 7 object characterizations (the eazy-py and Lephare characterizations for both COSMOS2020 catalogs).  }
	\label{tab:count_sumg}
	\begin{tabular}{lcrrrrr}
		\hline
		Catalog	& Characterization	& Total \sumg\ 	& Counterparts  & Counterparts		& Counterparts which are	& Counterparts which are  \\
			   	&				& Candidates 	& in 1+ Other     & in 3+ Other   	& \sumg s in 1+	 Other	& \sumg s in 5+ Other	    \\
			   	&				& 			& Catalogs	 & Catalogs		& Characterizations	 	& Characterizations   \\
		\hline
		\hline
		DR1 	  & EAZY+FAST	& 23	& 13	&   8  &   3   & 0  	\\
		DR3 	  &  EAZY+FAST 	&   6	&   4	&   2  &   0   & 0  	\\
		C15 	  & LePhare		&   4	&   2	&   2  &   1   & 0  	\\
		C20C &  eazy-py	    	& 45	& 45	& 39  & 14   & 0 	\\
		C20C &  LePhare		&   4	&   4	&   3  &   3   & 0  	\\
		C20F &  eazy-py	     	& 49	& 46	& 41  & 12   & 0  	\\
		C20F &  LePhare	     	& 10	& 10	&   8  &   2   & 0  	\\
	\end{tabular}
\end{table*}
%------------------------------------------

%------------------------------------------
%------------------------------------------

%------------------------------------------
%------------------------------------------

\clearpage

\section{Implications for the Number Densities of Massive Galaxies}\label{Sec:Disc}

\subsection{Number Density Corrections}

No \sumg\ candidates have $\lvert$~\zspec-\zphot~$\rvert<0.5$ in our primary selection catalogs, which appears to be due to difficulty in accurately modeling galaxies with very red SEDs.
Such SEDs have degeneracies between redshift, dust attenuation, AGN activity, and emission line strength which are not consistently and reliably fit by the different modeling codes tested herein.
This 0/16 success rate illustrates the extreme difficulty in accurately identifying such massive galaxies at these early epochs from ultraviolet, optical, and near-IR photometry alone.
To test the significance of this result, we assume that the 34 total high-quality \sumg\ candidates from the parent catalogs consist of a range of `true' \sumg s (\ie\ $3<z<4$ and \logM ~$>11.7$) from $N_{\rm real}$=0 to $N_{\rm real}$=34.
For each value of $N_{\rm real}$ we then randomly select 16 objects 1000 times to assess how often zero are real.
We confirm $N_{\rm real}<2$ at $1\sigma$ significance and $N_{\rm real}<8$ at $3\sigma$ significance. 
This implies that photometric catalogs overestimate the number of \sumg s by a factor $>4$ (99\% confidence) and perhaps by a factor $>17$ (68\% confidence).
A similar analysis of spectroscopically targeted MAGAZ3NE UMGs shows a much smaller correction of around $\lesssim10\%$, suggesting that such a correction is only significant for the most massive and/or the reddest candidates.

To visualize these offsets, we calculate number densities based on the average from each of the seven COSMOS catalogs and SED modeling variations considered in Section~\ref{Sec:AnComp}.
A comparison of the number density of UMG (\logM~$>11.0$) and \sumg\ (\logM~$>11.7$) candidates from the nominal cuts to each catalog ($3<$~\zphot~$<4$, $m_K<23.4$) show variations of up to $\sim 0.5$~dex from the values reported in the literature \citep{Muzzin2013,Marsan2022}.
This is understandable, as some targets were removed from these studies due to \eg\ photometric contamination from neighbors, proximity to stars or other bad regions, etc.
As such this can be thought of as an optimistic upper limit on the number densities of such targets.
We take the average of these values as number density measurements and the asymmetric standard deviations as the uncertainties.

\begin{figure*}
    \includegraphics[width=\textwidth, trim=0in 0in 0in 0in]{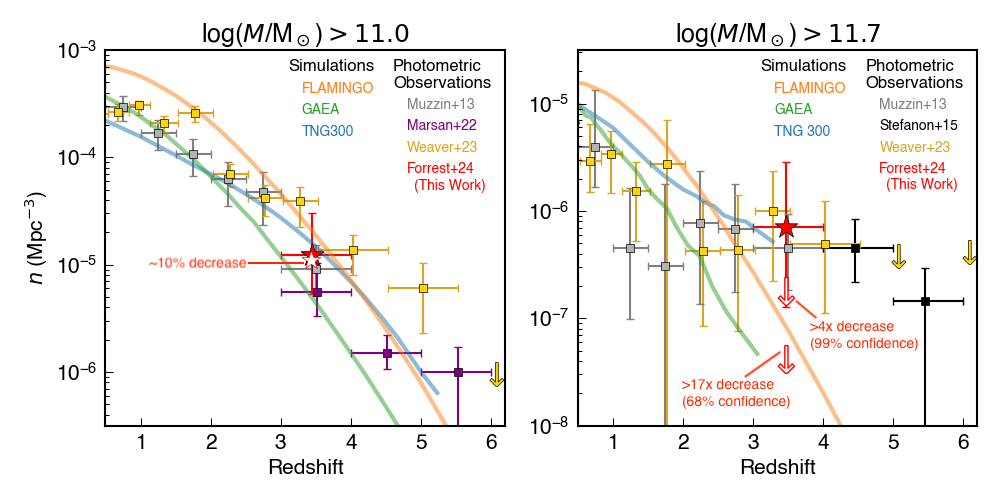}
    \caption{The number density of massive galaxy candidates in the COSMOS field. 	\textbf{Left:} Literature values for galaxies with \logM~$>11$ from \citet{Muzzin2013}, \citet{Marsan2022}, and \citet{Weaver2022} are shown as gray, purple, and gold squares, respectively.
        Values from the FLAMINGO, GAEA, and TNG300 simulations convolved with a mass uncertainty of 0.25~dex are shown as orange, green, and blue lines, which truncate when the number of galaxies above the mass threshold drops below ten.
	The filled red star is the average number density from the seven COSMOS catalogs and SED modeling characterizations considered in Section~\ref{Sec:AnComp}, while the unfilled red star on the left panel shows the correction to this number density for galaxies with \logM~$>11.0$ based on MAGAZ3NE spectroscopic success rate, a correction of $\sim10\%$.
	\textbf{Right:} Literature values for galaxies with \logM~$>11.7$ from \citet{Muzzin2013}, \citet{Stefanon2015}, and \citet{Weaver2022} are shown as gray, black, and gold squares, respectively.
    The unfilled red arrows show the spectroscopic correction for galaxies with \logM~$>11.7$, which suggest a decrease of at least $4\times$ (99\% confidence) and possibly more than $17\times$ (68\% confidence).
    }
    \label{fig:numden}
\end{figure*}

These values are shown as filled red stars in Figure~\ref{fig:numden} and agree well with other literature values.
The effects of the correction factors derived above are shown as unfilled red symbols - a star for the UMGs and arrows (upper limits) for the \sumg s.

%------------------------------------------
%------------------------------------------

\subsection{Comparison with Simulations}

The high-mass end of the GSMF has long shown discrepancies between observations and simulations at high redshifts, implying either inaccuracies or systematics when inferring the physical properties of such galaxies in observations or difficulties in building sufficient stellar mass at early times in simulations \citep[or both; \eg][]{Steinhardt2016, Sherman2020a}.

Our results suggest that observational-related uncertainties play a significant role.
It is also worth noting that such massive galaxies are very rare at early times, and thus require large volumes in both observations and simulations to include these objects in statistically meaningful numbers.
To this end, we compare the number densities of UMGs from photometric catalogs to those in the newest generation of models with large volumes, including:
1) the TNG300 hydrodynamical simulation of the IllustrisTNG project \citep{Springel2018, Pillepich2018, Nelson2018, Nelson2019}, with a volume of \mbox{(205 Mpc/h)$^3$}  ($h=0.6774$),
2) the FLAMINGO suite of simulations \citep{Schaye2023, Kugel2023}, considering the largest volume box, L2p8\_m9, \mbox{(2.8 Gpc)$^3$} in volume ($h=0.681$), and
3) the GAEA semi-analytic model \citep{deLucia2024}, which builds upon the \mbox{(500 Mpc/h)$^3$} box from the Millenium dark matter only simulation \citep[$h=0.73$;][]{Springel2005}.

We consider predictions over the redshift range $0.5<z<6$, and convolve the intrinsic stellar mass of each galaxy with a Gaussian of center unity and standard deviation 0.25~dex to simulate the systematic uncertainties from different stellar population modeling codes \citep[\eg][]{Mobasher2015, Wang2023a}.
We consider stellar mass inside a projected aperture of 30~kpc for FLAMINGO and within twice the half-stellar mass radius for TNG300, while stellar masses from GAEA are `total' stellar masses.
Comparison of these choices and other factors contributing to differences between simulations and observations are presented in Appendix~\ref{App:Sim}.
We note however that there will still be discrepancies between these approximations and a full photometric extraction and catalog construction of the simulation data emulating the exact methods used for observations.
Such a detailed comparison is beyond the scope of this work.

The number density comparison is shown in Figure~\ref{fig:numden}.
Broadly speaking, the agreement in number densities of UMGs between observations and simulations is good.
Number densities from different observational works are consistent with each other out to $z\lesssim4$, and diverge when moving to higher redshifts.
TNG300 replicates the observational number densities over most of the $0.5<z<6$ redshift range considered, with a slight underprediction at low redshifts.
The growth of number densities over time from FLAMINGO and GAEA are similar, but there is a normalization offset of a few times in number density over all redshifts.

There is more uncertainty in the number densities of \sumg s, much of which is driven by observations.
After decreasing out to $z\sim2$, observational number densities appear to flatten, which may be consistent with measurements being dominated by Eddington bias and Poissonian statistics.
The lack of spectroscopically-confirmed \sumg s in this work suggests that the number densities of such objects does in fact continue to decrease to higher redshifts.
While simulations do show a continued dropoff in \sumg\ number density at $z>2$, even the large volumes of the GAEA and TNG300 simulations do not predict statistically meaningful numbers of \sumg s at high redshifts.
The volume of the largest FLAMINGO simulation allows a prediction of \sumg\ number densities below the observed values, but still within $\lesssim2\sigma$ due to the large observational uncertainties, and also in good agreement with the observational correction derived from this work.
Spectroscopic followup of larger and more complete samples of photometric \sumg\ candidates is necessary to better constrain whether a discrepancy between simulations and observations is significant.
With a larger study it could be quite possible that this tension disappears altogether.

\subsection{Application to Massive Photometric High-Redshift Candidates}

We have attempted to use a combination of spectroscopy and photometry to generate SED models for use with photometric redshift fitting codes for similar galaxies.
These efforts have shown only limited success, implying that the problem may not be simply a lack of sufficiently similar models, but also the application of priors which do not accurately reflect the existence of this population of galaxies.
Additionally, a stack of Keck/NIRES spectra of lower redshift galaxies with extreme dust obscuration and strong rest-frame optical emission lines \citep[hot DOGs;][]{Finnerty2020} shows that such galaxies can have degenerate JWST NIRCam colors with red, high redshift galaxies \citep{McKinney2023}.
As a result, is is possible that such misidentifications from photometric data occur over a range of redshifts.

Indeed, large numbers of candidate massive galaxies at surprisingly early times have been identified in JWST data, \citep[\eg][]{Naidu2022a, Naidu2022, Labbe2023, Barro2024, Kocevski2024}.
A handful of these have been spectroscopically confirmed as early, massive galaxies \citep[\eg][]{Carnall2023, Haro2023a, Fujimoto2023a}.
However, populations of red galaxies with significant AGN flux contributions have also been confirmed with JWST and longer wavelength data, some of which have significantly higher photometric redshifts than later confirmed \citep[\eg][]{Haro2023, Fujimoto2023a, Meyer2023}.
Further JWST/NIRSpec follow-up of these candidates will shed light on the hypothesis that a significant fraction of red, massive high-redshift candidates are in fact lower redshift galaxies and/or less massive due to inaccurate photometric accounting of dust and AGN activity.

\clearpage

\section{Conclusions}\label{Sec:Conc}

Deep, multi-passband photometric catalogs point to the existence of a population of galaxies with redder colors and greater stellar masses than the most massive galaxies which have previously been spectroscopically confirmed at $3<z<4$.
In this work, the Massive Ancient Galaxies at $z>3$ Near-Infrared (MAGAZ3NE) Survey spectroscopically targeted 16 such galaxy candidates with \zphot~$\gtrsim3$ and \logM~$>11.7$ (which we term \sumg s) across three fields with some of the most extensive photometric data and high-quality photometric catalogs (UltraVISTA-COSMOS, VIDEO-CDFS, VIDEO-XMM).

In sharp contrast to slightly lower mass galaxies from these same catalogs, which show excellent agreement between spectroscopic and photometric redshifts, none of the \sumg\ candidates has a redshift \mbox{$\lvert$~\zspec-\zphot~$\rvert<0.5$}.
Three galaxies show broad emission lines at $z>3$ indicating Type I AGN contamination of the photometry biasing the photometric estimates of redshift and stellar mass.
Half of the sample (9/16) are confirmed to be lower redshift interlopers at $2.0<z<2.7$ (8 objects) and $z\sim1.2$ (1 object) but are still very massive.
The remaining four targets do not have confirmed spectroscopic redshifts.

A comparison of catalogs in the COSMOS field suggests that none of the various construction methods or SED fitting methodologies used in these catalogs characterize the redshifts and stellar masses of these galaxies accurately.
The underlying issue appears to be degeneracies among redshift, AGN contribution, and dust attenuation for objects with such red observed colors, which SED modeling templates have trouble disentangling.

We use the results of our spectroscopic observations to revise the number density of \sumg s seen in photometric surveys.
This correction is a reduction by a factor of $>4\times$ (99\% confidence) and likely $>17\times$ (68\% confidence).
A comparison to some of the newest and largest volume models shows that this reduction reduces the severity of the discrepancy between simulations and observations at $3<z<4$.
These misidentifications thus contribute substantially, and possibly entirely, to the discrepancy between observations and simulations in the number densities of the most massive galaxies at $3<z<4$, reducing the observed apparent flattening in number densities at high redshifts for these very massive objects.

The apparent failure of photometric redshift fitting codes to accurately identify these objects from fields with some of the most extensive and deepest photometry available suggests that improved optical and near-infrared photometry alone is unlikely to solve this problem over wide fields.
While concentrated surveys with medium- and narrow-band photometry may reduce the severity of this issue, as could further imaging from facilities such as ALMA and JWST, spectroscopic campaigns targeting such objects are the only way to truly quantify the scale of this effect.

\section*{Acknowledgments}

The authors wish to recognize and acknowledge the very significant cultural role and reverence that the summit of Maunakea has always had within the indigenous Hawaiian community.  We are most fortunate to have the opportunity to conduct observations from this mountain.

This research was supported by the International Space Science Institute (ISSI) in Bern, through ISSI International Team project ``Understanding the evolution and transitioning of distant proto-clusters into clusters".  We are grateful for the support of ISSI and the use of their facilities.
We also gratefully acknowledge the Lorentz Center in Leiden (NL) for facilitating discussions on this project. 

The FLAMINGO simulation used the DiRAC@Durham
facility managed by the Institute for Computational Cosmology on
behalf of the Science and Technology Facilities Council (STFC)
Distributed Research Utilising Advanced Computing (DiRAC) High
Performance Computing Facility (www.dirac.ac.uk). The equipment
was funded by BEIS capital funding via STFC capital grants
ST/K00042X/1, ST/P002293/1, ST/R002371/1, and ST/S002502/1,
Durham University and STFC operations grant ST/R000832/1.

GW gratefully acknowledges support from the National Science Foundation through grant AST-2205189 and from HST program number GO-16300.
DN acknowledges funding from the Deutsche Forschungsgemeinschaft (DFG) through an Emmy Noether Research Group (grant number NE 2441/1-1).

BF thanks Brian C. Lemaux and Ian Smail, as well as the anonymous referee, for helpful comments on the manuscript.

This work has relied heavily upon code developed by other people, for which we are quite thankful.
\software{
Astropy \citep{Astropy2013,Astropy2018,Astropy2022},
Bagpipes \citep{Carnall2018},
EAZY \citep{Brammer2008},
FAST \citep{Kriek2009},
FAST++ \citep{Schreiber2018a},
IPython \citep{Perez2007},
LePhare \citep{Arnouts1999, Ilbert2006}
Matplotlib \citep{Hunter2007},
NumPy \citep{Oliphant2006}
}

\bibliography{library} 

%%%%%%%%%%%%%%%%%%%%%%%%%%%%%%%%%%%%%%%%%%%%%%%%%%

\appendix
\renewcommand\thefigure{\thesection.\arabic{figure}}

\section{Additional \sumg\ Candidate Information}
\setcounter{figure}{0}

\begin{figure*}
	\includegraphics[width=\textwidth, trim=0in 4.4in 3.75in 0in]{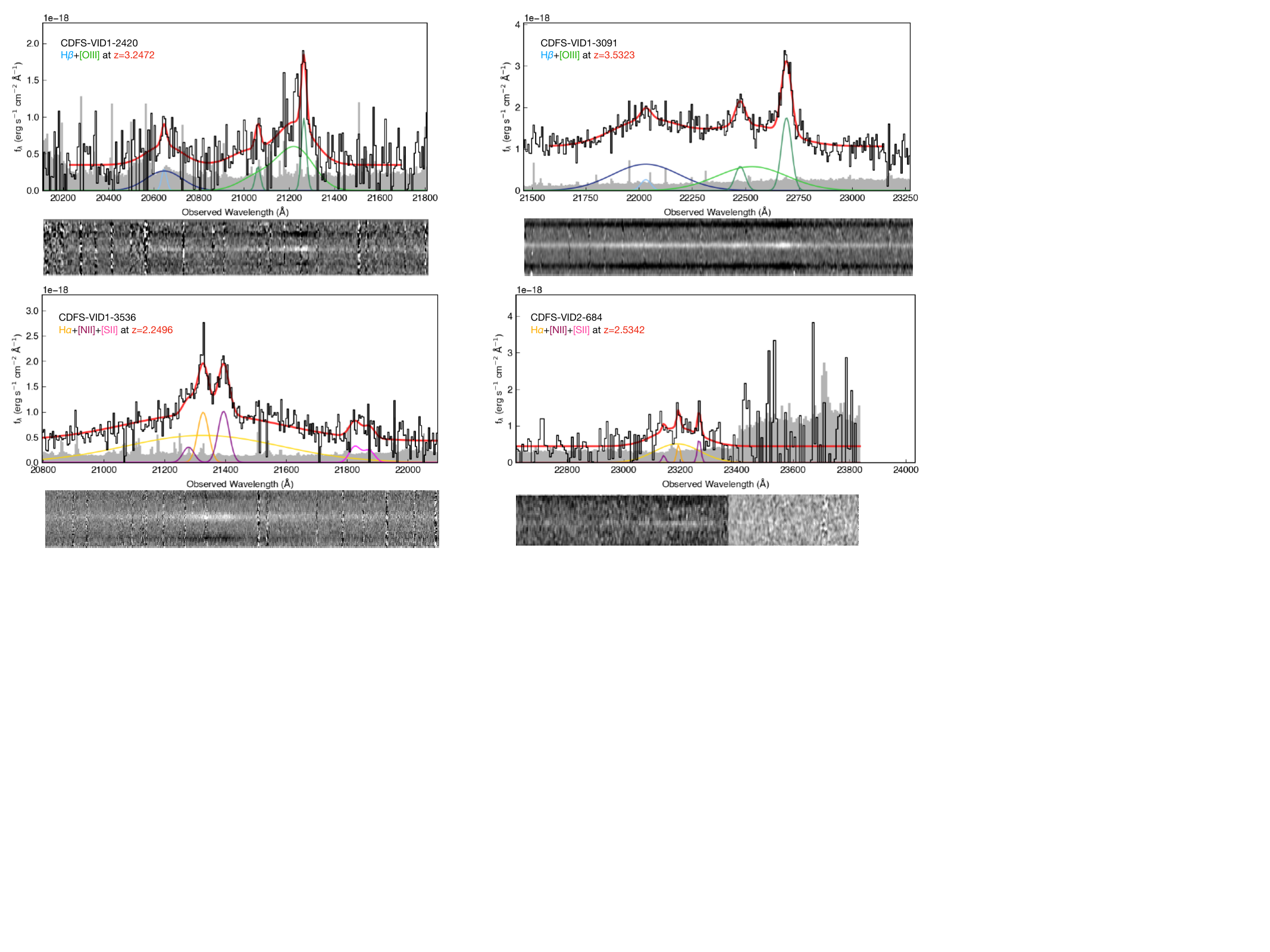}
    \caption{Line complex fitting for obtaining spectroscopic redshifts of candidate \sumg s in the VIDEO-CDFS field with the same layout as Figure~\ref{fig:speclines}. For objects with confirmed redshifts, the region of the 1D and 2D spectrum with the confirming lines and fitted model are shown. The overall model is shown in red, while individual line components are shown in different colors (\Hbeta\ in blue, \OIII\ in green, \Halpha\ in orange, \NII\ in purple, and \SII\ in magenta). For objects without confirmed redshifts, the entire wavelength range probed by $H$- and $K$-band spectroscopy is shown. Some targets were observed with multiple masks which were sensitive to different wavelengths; in such cases regions covered by fewer than the total number of masks are displayed as a lighter shade of gray.}
    \label{fig:speclines_CDFS}
\end{figure*}

\begin{figure*}
	\includegraphics[width=\textwidth, trim=0in 1.5in 0.4in 0in]{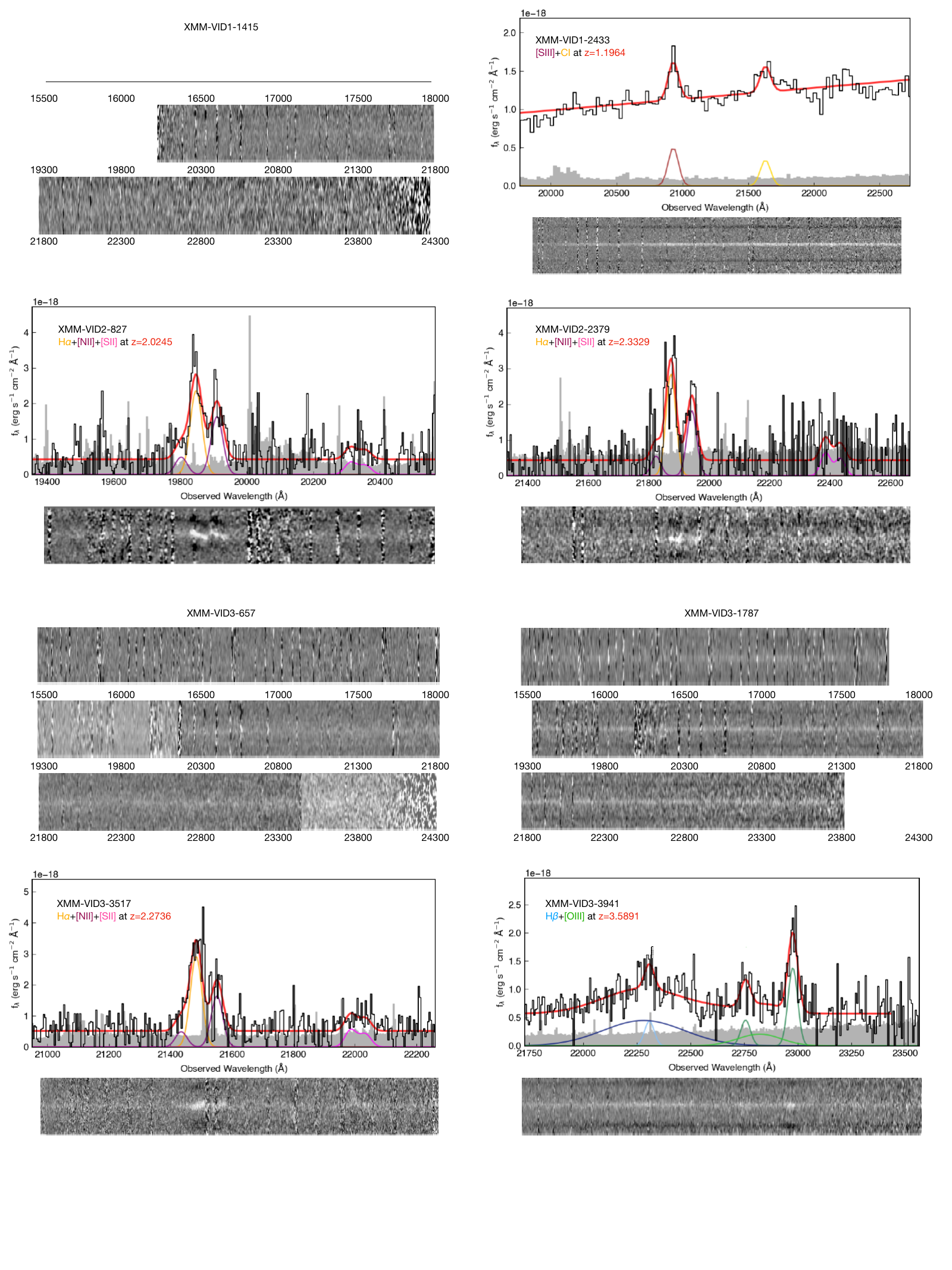}
    \caption{Line complex fitting for obtaining spectroscopic redshifts of candidate \sumg s in the VIDEO-XMM field. Same layout as Figure~\ref{fig:speclines}.}
    \label{fig:speclines_XMM}
\end{figure*}

Similar to Figure~\ref{fig:speclines}, the 2D and 1D spectra with emission line modeling for \sumg\ candidates in the CDFS and XMM fields are shown in Figures~\ref{fig:speclines_CDFS} \& \ref{fig:speclines_XMM}, respectively.
Table~\ref{tab:photprop} reports the coordinates, observing conditions, and properties of all targeted MAGAZ3NE UMG and \sumg\ candidates.

%------------------------------------------
%------------------------------------------

%------------------------------------------
%------------------------------------------

\begin{table*}
	\centering
	\caption{The list of photometric UMG and \sumg\ candidates targeted by the MAGAZ3NE survey with Keck/MOSFIRE and Keck/NIRES. Publications listed are:
	a)~\citet{Forrest2020a},
	b)~\citet{Forrest2020b},
	c)~\citet{Forrest2022},
	d)~This Work,
	e)~\citet{Marsan2015},
	f)~\citet{Marsan2017},
	g)~\citet{Saracco2020},
	h)~\citet{McConachie2022},
	i)~McConachie+2024, submitted,
	j)~Chang+2024, in prep,
	k)~Stawinski+2024, in prep.
	l)~\citet{Shen2021}}
	\label{tab:photprop}
	\begin{tabular}{lrrccrccrrc} 
		\hline
		Galaxy & R.A. & Dec. & $t_{\rm exp, K (H)}$ & Seeing$_{\rm K (H)}$  & $m_K$  & \zphot & \zspec & $M^*_{\rm phot} $ & $M^*_{\rm spec}$ & Refs.\\
			& hh mm ss & dd mm ss & (min) & ($\arcsec$) & & & & log(M$_\odot$) & log(M$_\odot$) & \\
		\hline
		XMM-VID1-896		&  02  15  36.408  	& -04  52  19.33	&120		& 0.65	& 22.82	& 4.133	& --		& 11.05	& --	& k	\\
		XMM-VID1-1126	&  02  16  46.821  	& -04  13  46.30	& 72		& 0.75	& 21.65	& 3.740	& 3.495	& 11.64	& 11.59	& d	\\
		XMM-VID1-1415	&  02  19  01.864  	& -05  11  14.53	& 66		& 0.70 	& 21.34	& 3.016	& --		& 11.84	& -- 	& d	\\
		XMM-VID1-2075	&  02  17  32.863  	& -05  28  57.39	& 84		& 0.70	& 20.80	& 3.474	& 3.452	& 11.50	& 11.52	& b, c	\\
		XMM-VID1-2399	&  02  18  01.830 	& -05  05  11.83	&312		& 0.85	& 22.05	& 3.680	& 3.580	& 11.14	& 11.02	& b	\\
		XMM-VID1-2433	&  02  17  05.599  	& -04  20  32.30 	& 260 (=) 	& 0.80 	& 20.66 & 3.385 & 1.196	& 11.77	& 10.18	& d	\\
		XMM-VID1-2761	&  02  19  21.082  	& -04  30  10.74	& 312	& 0.84	& 21.75	& 3.596	& --		& 11.43	& --	& d	\\	
		\hline
		XMM-VID2-827		&  02  23  43.360  	& -04  50  33.85	& 72		& 0.66	& 21.41	& 3.033	& 2.024	& 11.88 	& 10.74	 	& d	\\
		XMM-VID2-2379	&   02  20  44.230  	& -05  06  27.35	& 96		& 0.64	& 21.71	& 3.317	& 2.333	& 11.77	& 10.75	 	& d	\\
		\hline
		XMM-VID3-657		&  02  25  40.887  	& -04  16  08.95	&354 (96)	& 0.69 (0.75) & 21.66 & 3.683	& --		& 11.88	& --	& d	\\
		XMM-VID3-1120	&  02  27  10.098  	& -04  34  44.99	&180	 (376) & 0.60 (1.2) & 21.17 & 3.445	& 3.492	& 11.48	& 11.47	& a, b, c	\\
		XMM-VID3-1638	&  02  24  50.993  	& -05  10  31.74	& 60		& 0.55 	& 22.42	& 4.059	& 3.824	& 11.23	& 11.18	& k	\\
		XMM-VID3-1787	&  02  26  22.333 	& -04  42  56.63	&114 (80)	& 0.60 (0.85) & 21.15 & 3.051	& --		& 11.99 	& --		& d	\\
		XMM-VID3-2293	&  02  26  34.076 	& -04  22  19.37	& 48		& 0.85	& 21.35	& 3.075	& 3.313	& 11.56	& 11.57	& b, l 	\\
		XMM-VID3-2457	&  02  26  56.043 	& -04  32  11.64	&180		& 0.60	& 21.52	& 3.513	& 3.489	& 11.23	& 11.26	& b, c	\\
		XMM-VID3-3517	&  02  24  12.920  	& -04  58  53.58	& 296 (128) & 0.85 (0.85) & 21.42 & 3.267 & 2.274	& 11.78	& 11.35	 	& d	\\
		XMM-VID3-3941	&  02  24  13.691  	& -04  40  10.94	&180		& 1.1		& 21.33	& 3.042	& 3.590	& 11.79	& 11.53	 	& b, d	\\
		\hline
		\hline
		CDFS-VID1-2420	&  03  31  31.881  	& -28  04  10.32	&132		& 0.74	& 21.69	& 3.784	& 3.245	& 12.07	& 11.91	  	& d	\\
		CDFS-VID1-3091	&  03  28  00.346  	& -27  31  50.77	&114	 (40) & 0.60 (0.70) & 20.59 & 2.995	& 3.532	& 11.85	& 11.67	  	& d	\\
		CDFS-VID1-3536	&  03  27  39.060 	& -28  02  09.25 	&204		& 0.67	& 21.44	& 3.274	& 2.25	& 11.87	& 11.42	   	& d	\\
		\hline
		CDFS-VID2-684	&  03  30  27.064  	& -28  21  13.22	&210 	& 1.0		& 21.67	& 3.323	& 2.533	& 11.80	& 11.63	  	& d	\\
		\hline
		\hline
		COS-DR1-70455	&  09  58  02.923   	& 01  57  55.62		&312		& 0.58	& 21.85	& 3.149	& 2.655	& 11.75	& 11.33	 	& d	\\
		COS-DR1-79837	&  10  02  32.244   	& 02  06  32.18		&120	 (216) & 0.74 (0.85) & 21.10 & 3.318	& --		& 11.99	& --		& d	\\
		COS-DR1-99209	&  10  02  56.366   	& 02  24  06.00		& 54 (72)	& 0.90 (1.2) & 20.90	& 3.071	& 2.983	& 11.16	& 11.22		&  c	\\
		COS-DR1-113684	&  09  59  43.891   	& 02  07  24.28		&204		& 0.75	& 21.39	& 3.473	& 3.831	& 11.23	& 11.20	& b, c	\\
		COS-DR1-130749	&  10  01  22.382   	& 02  20  02.73		&144		& 0.85	& 21.54	& 2.984	& --		& 11.29	& --		& d	\\
		COS-DR1-187247	&  10  02  49.205   	& 02  32  55.76		&162		& 0.65	& 22.04	& 3.314	& --		& 11.46	& --		& d	\\
		COS-DR1-209435	&  10  00  04.793   	& 02  30  45.28		&252		& 0.68	& 22.03	& 3.769	& 2.481	& 11.98	& 11.55	& d, j		\\
		COS-DR1-233254	&  10  01  	03.578	& 02  48  10.22		&69		& 0.77	& 21.10	& 2.880	& 2.059	& 11.78	& 10.78		& d	\\
		COS-DR1-258857	&  09  58  18.120  	& 02  45  36.20		&120		& 0.70	& 21.66	& 3.269	& --		& 11.39	& --		& d	\\
		\hline
		COS-DR3-9705	&  10  00  13.481   	& 01  37  04.76		& 84		& 0.65	& 21.92	& 3.108	& --		& 11.47	& --		& d	\\
		COS-DR3-61518	&  10  02  05.695    	& 02  13  50.12		&180		& 0.75	& 22.26	& 3.233	& 3.367	& 11.33	& 11.31	& d	\\
		COS-DR3-84674	&  10  00  43.759   	& 02  10  28.68 	&0 (280)	& -- (0.9)	& 21.28	& 3.067	& 3.009	& 11.23	& 11.25	& b, c	\\
		COS-DR3-111740	&  09  58  53.746   	& 02  10  38.63		&156		& 1.3		& 21.10	& 3.149	& 2.799	& 11.11	& 10.98	& b, c	\\
		COS-DR3-131150	&  10  01  46.001   	& 02  29  49.13		&372		& 0.75	& 22.53	& 3.031	& 3.112	& 10.82	& 10.84	& i	\\
		COS-DR3-131925	&  10  01  42.662   	& 02  30  20.87 	&372		& 0.75	& 20.96	& 3.191	& 3.139	& 11.11	& 10.27	& b, i		\\
		COS-DR3-133914	&  10  01  58.922	& 02  32  27.98		&372		& 0.75	& 23.24	& 3.129	& 3.127	& 10.52	& 10.52	& i	\\
		COS-DR3-134032	&  10  01  59.155   	& 02  32  29.84 	&372		& 0.75	& 22.44	& 2.918	& 3.137	& 10.97	& 10.97	& i	\\
		COS-DR3-160748	&  10  00  27.811   	& 02  33  49.23		& 60		& 0.61	& 20.26	& 3.318	& 3.352	& 11.50	& 11.46	& b, e, f, g, h, i	\\
		COS-DR3-179370	&  09  59  24.393   	& 02  25  36.51 	&288		& 0.78 	& 22.14	& 3.629	& 3.367	& 11.32	& 11.37	& b, e, f, h	, i\\
		COS-DR3-186449	&  09  58  56.899  	& 02  32  58.01		& 96		& 1.1		& 21.54	& 3.675	& --		& 11.10	& --		& d	\\
		COS-DR3-195616	&  09  59  24.590   	& 02  43  10.13		&132		& 0.85	& 21.64	& 3.067	& 3.255	& 11.26	& 11.31	& b, j		\\
		COS-DR3-201999	&  09  57  48.578   	& 01  39  57.69		&120	 (352) & 0.75 (1.0) & 21.00 & 3.108	& 3.131	& 11.37	& 11.40	& b, c	\\
		COS-DR3-202019	&  09  57  48.516   	& 01  40  05.06		&120	 (352) & 0.75 (1.0) & 20.79 & 3.108	& 3.133	& 11.61	& 11.67	& b, c	\\
		COS-DR3-208070	&  09  57  53.815   	& 01  51  57.52		&132		& 0.75 	& 21.70	& 3.448	& 3.491	& 11.24	& 11.26	& b	\\
		COS-DR3-226441	&  09  57  36.156   	& 02  28  10.63		&288		& 0.75	& 21.65	& 3.275	& 3.245	& 11.10	& 11.02	& b	\\
	\end{tabular}
\end{table*}

%%%%%%%%%%%%%%%%%%%%%%%%%%%%%%%%%%%%%%%%%%%%%%%%%%

\section{Simulation Considerations}\label{App:Sim}
\setcounter{figure}{0}

\begin{figure*}
    \centering
    \includegraphics[width=0.75\textwidth, trim=0in 2in 0in 0in]{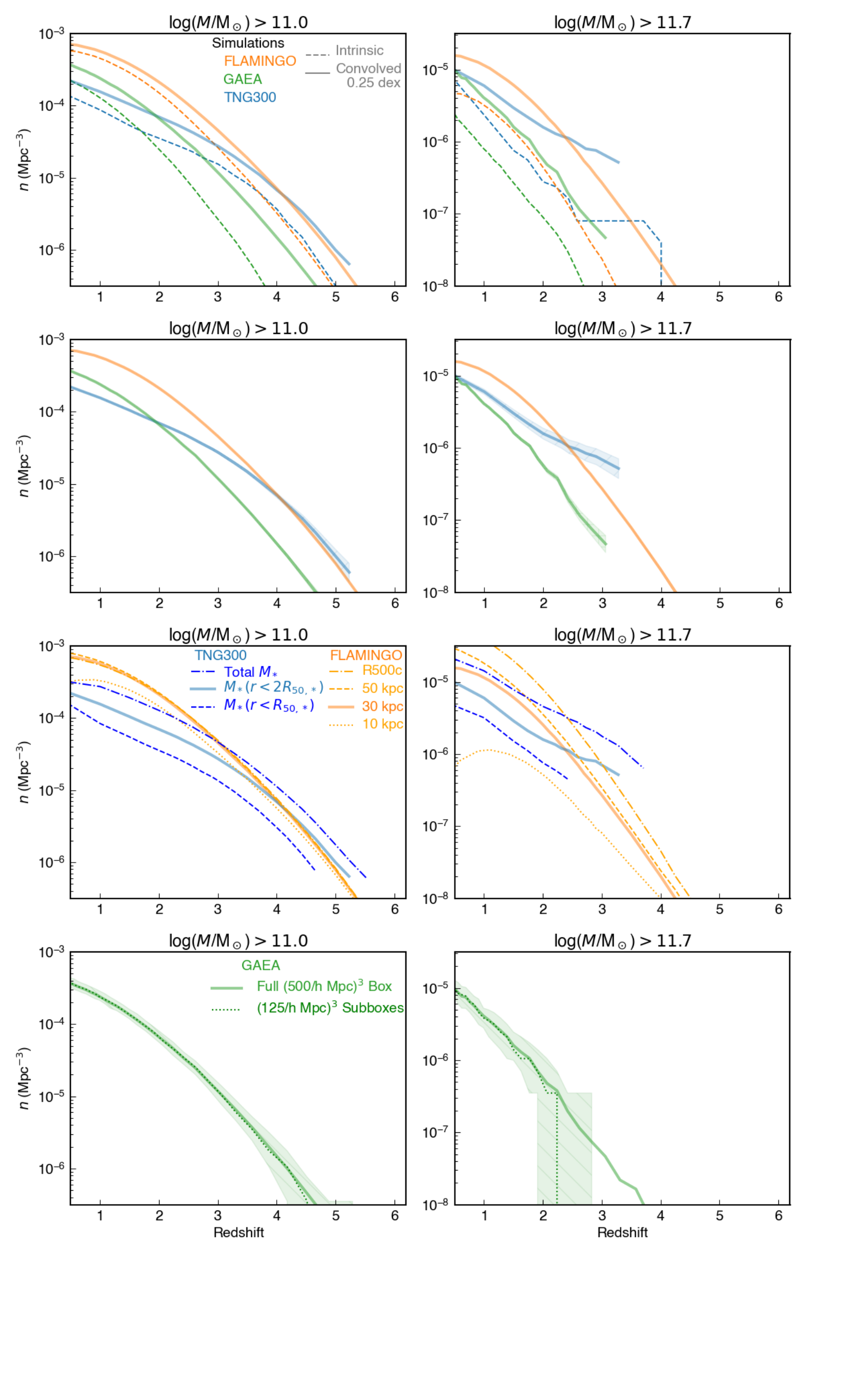}
    \caption{The effects of different simulation comparison uncertainties. Left panels show UMGs and right panels show \sumg s. FLAMINGO, GAEA, and TNG300 are shown in orange, green, and blue, respectively.
    \textbf{Top row:} The intrinsic number densities from simulations (dashed lines) compared to the values after resampling each stellar mass using a Gaussian of width 0.25~dex (solid lines).
    \textbf{Second row:} The scale of Poissonian uncertainties in each simulation (shaded areas). Values are truncated when the number of objects in the simulation passing the mass cut drops below ten.
    \textbf{Third row:} Number densities resulting from different aperture selection. The FLAMINGO comparison includes lines for three projected aperture radii in physical units, as well as a spherical aperture with radius $R_{500c}$, where the density is 500 times the critical density of the Universe. The TNG300 comparison includes lines considering the total stellar mass in a subhalo, all stellar mass within the stellar half-mass radius, and all stellar mass within twice the stellar half-mass radius.
    \textbf{Bottom row:} Number densities in the entire GAEA (500/h Mpc)$^3$) box (solid line), and those from (125/h Mpc)$^3$) subboxes. The median is shown as a dotted line, with shading between the 16$^{\rm th}$ and 84$^{\rm th}$ percentile values. Values are shown even when the number of galaxies in a bin are less than ten.
    }
    \label{fig:simApp}
\end{figure*}

Number densities from the FLAMINGO, GAEA, and TNG300 simulations can be derived in various ways.
Additionally, simulation data require application of observational effects in an attempt to perform fair comparisons.
In Figure~\ref{fig:simApp} we show comparisons of these choices and sources of uncertainty.

%------------------------------------------
%------------------------------------------

\end{document}